# Nanosecond electro-optics of nematic liquid crystal with negative dielectric anisotropy

Volodymyr Borshch,[1] Sergij V. Shiyanovskii,[1] Bing-Xiang Li,[1,2] and Oleg D. Lavrentovich[1,*]

[1]*Liquid Crystal Institute and Chemical Physics Interdisciplinary Program, Kent State University, Kent, OH, 44242, USA*

[2]*College of Electronic and Information Engineering, Nanjing University of Aeronautics and Astronautics, Nanjing 210016, China*

**Abstract.** We study a nanosecond electro-optic response of a nematic liquid crystal in a geometry where an applied electric field **E** modifies the tensor order parameter but does not change the orientation of the optic axis (director $\hat{\mathbf{N}}$). We use a nematic with negative dielectric anisotropy with the electric field applied perpendicularly to $\hat{\mathbf{N}}$. The field changes the dielectric tensor at optical frequencies (optic tensor) due to the following mechanisms: (a) nanosecond creation of the biaxial orientational order; (b) uniaxial modification of the orientational order that occurs over timescales of tens of nanoseconds, and (c) the quenching of director fluctuations with a wide range of characteristic times up to milliseconds. We develop a model to describe the dynamics of all three mechanisms. We design the experimental conditions to selectively suppress the contributions from the quenching of director fluctuations (c) and from the biaxial order effect (a) and thus, separate the contributions of the three mechanisms in the electro-optic response. As a result, the experimental data can be well fitted with the model parameters. The analysis provides a rather detailed physical picture of how the liquid crystal responds to a strong electric field on a timescale of nanoseconds. This work provides a useful guidance in the current search of the biaxial nematic phase. Namely, the temperature dependence of the biaxial susceptibility allows one to estimate the temperature of the potential uniaxial-to-biaxial phase transition. An analysis of the quenching of director fluctuations indicates that on a timescale of nanoseconds, the classic model with constant viscoelastic material parameters might reach its limit of validity. The effect of nanosecond electric modification of the order parameter (NEMOP) can be used in applications in which one needs to achieve ultrafast (nanosecond) changes of optical characteristics, such as birefringence.

---

* Electronic mail: olavrent@kent.edu

#### I. INTRODUCTION

The uniqueness of nematic liquid crystal (NLC) materials is defined by the long-range orientational order of their constituent molecules, which have anisometric shape, permanent and induced dipoles [1]. The average orientation of NLC molecules in a certain point in space, described by the radius-vector $\mathbf{r}$, is called the director $\hat{\mathbf{N}}(\mathbf{r})$, which coincides with its optic axis. Director orientation can vary from point to point in space or fluctuate in time.

Anisotropic optic and dielectric properties of NLCs, namely, birefringence $\Delta n = n_e - n_o$, where $n_e$ and $n_o$ are the extraordinary and ordinary refractive indices, respectively, and dielectric anisotropy $\Delta \varepsilon = \varepsilon_\parallel - \varepsilon_\perp$, with $\varepsilon_\parallel$ measured along and $\varepsilon_\perp$ perpendicular to the optic axis, enabled wide range of electro-optic applications. Traditional electro-optic applications of NLCs are based on field-induced reorientation of $\hat{\mathbf{N}}$, known as the Frederiks effect. For $\Delta \varepsilon > 0$, the director realigns parallel to an applied electric field $\mathbf{E}$, while for $\Delta \varepsilon < 0$, it realigns perpendicularly to the field. The characteristic switch-on time is $\tau_{on}^F = \gamma_1 / \varepsilon_0 |\Delta \varepsilon| E^2$, where $\gamma_1$ is the rotational viscosity, and $\varepsilon_0$ is the electric constant. The switch-off time $\tau_{off}^F = \gamma_1 d^2 / K \pi^2$ is typically slower, in the range of milliseconds, being determined by the elastic constant $K$ of the NLC (typically 10 pN) and the cell thickness $d$ (typically 5 µm).

An electro-optic response of the LC, however, can be triggered without director realignment, as it suffices to modify the tensorial order parameter (OP) without altering its orientation [2-12]. An important feature of this approach is that the OP modifications of both uniaxial and biaxial nature take place at the molecular scale and, thus, are very fast (nanoseconds and tens of nanoseconds [12, 13]) for both field-on and field-off driving. For this reason, it is convenient to call the pure OPs-related phenomenon a "nanosecond electric modification of the order parameters" effect, or the NEMOP effect. In addition to the modification of the OPs, the applied field also quenches the director fluctuations [1, 11, 14-25]. The later effect, being of macroscopic origin, is typically much slower, as determined by the length scale of fluctuative director distortions. Both the fundamental understanding and practical applications of NEMOP require one to separate the fast effects of NEMOP and the slow effects of director fluctuations. This problem and its solution represent one of the main focuses of the presented work.

In this work, we demonstrate how to separate the NEMOP effect and the dynamics of director fluctuations by choosing a particular geometry of light propagation through a cell filled with a planar NLC of a negative dielectric anisotropy. The electric field is applied perpendicularly to $\hat{\mathbf{N}}$. Section II presents a theoretical model of the dynamics of the uniaxial and biaxial modifications of the OP and the dynamics of director fluctuations in the electric field. It is shown that the contributions originating in the OP changes and in director fluctuations can be separated from each other by testing the cell under different angles of light incidence. Section III describes the experimental set-up to measure the field-induced optic response, which occurs at short timescales down to nanoseconds. Our approach allows one to separate the field-induced birefringence from parasitic effects, such as light scattering. Section IV describes the fitting of the experimental results with the proposed models. Section V discusses the physical mechanisms involved in the ultrafast electro-optic response of an NLC and utilization of the data in evaluating the likelihood of the appearance of a biaxial nematic phase in a field-free state.

## II. THEORY

Electro-optic processes could be considered using the free energy functional describing the NLC in the presence of an external electric field:

$$F = \int_V \left( f_{iso} + f_m + f_e + f_d \right) dV , \qquad (1)$$

where $f_{iso}$ is the free energy density of the isotropic phase for $E = 0$, $f_m = f_m(R_{jk})$ is the phenomenological microscopic free energy density written in the Landau formalism that depends on the scalar order parameters (OPs) $R_{jk}$, $f_e$ is the elastic free energy density due to distortions of $\hat{\mathbf{N}}$, and $f_d = -\frac{1}{2}\varepsilon_0 \mathbf{E}\boldsymbol{\varepsilon}\mathbf{E}$ is the anisotropic dielectric coupling energy density. The dielectric tensor $\boldsymbol{\varepsilon}$ depends on the OPs $R_{jk}$ and director fluctuations and can be represented as $\boldsymbol{\varepsilon}(R_{jk}, \hat{\mathbf{N}}) = \boldsymbol{\varepsilon}^{(0)}(R_{jk}^{(0)}, \hat{\mathbf{N}}_0) + \delta\boldsymbol{\varepsilon}^{(m)}(R_{jk}, \hat{\mathbf{N}}_0) + \delta\boldsymbol{\varepsilon}^{(fl)}(R_{jk}^{(0)}, \hat{\mathbf{N}})$, where $\boldsymbol{\varepsilon}^{(0)}(R_{jk}^{(0)}, \hat{\mathbf{N}}_0)$ is the field-independent tensor defined for a static and uniform (no fluctuations) director $\hat{\mathbf{N}}_0$, $\delta\boldsymbol{\varepsilon}^{(m)}$ is the field-induced modification associated with the OPs, and $\delta\boldsymbol{\varepsilon}^{(fl)}$ is the modification of the tensor caused by the director fluctuations $\delta\hat{\mathbf{N}}(\mathbf{r}) = \hat{\mathbf{N}}(\mathbf{r}) - \hat{\mathbf{N}}_0$, which depend on the applied electric field. We neglect higher order terms, such as coupling between the director fluctuations and field-induced changes in OPs. The terms containing $\delta\boldsymbol{\varepsilon}^{(m)}$ in the dielectric energy density $f_d$ define the effect of electrically-modified OPs. The term containing $\delta\boldsymbol{\varepsilon}^{(fl)}$ in $f_d$ influences the spectrum of director fluctuations.

## A. Dynamics of NEMOP effect

The orientational OPs can be described by the averaged Wigner $D$-functions $\langle D_{jk}^L \rangle$ [26-29], because $D_{jk}^L(\Omega)$ form a complete set of orthogonal functions of the Euler angles $\Omega = \{\omega_1, \omega_2, \omega_3\}$ [30]; $\Omega$ defines the molecular orientation through rotation $\mathbb{L} \xrightarrow{\Omega} \mathbb{M}$ from the laboratory frame $\mathbb{L}$ to the molecular frame $\mathbb{M}$. A set of OPs $\langle D_{jk}^L \rangle$, obtained by averaging with the single molecule orientational distribution function $f(\Omega)$, is complete and equivalent to

$$\langle D_{jk}^L \rangle = \int D_{jk}^L(\Omega) f(\Omega) \mathrm{d}\Omega. \qquad (2)$$

The nematic phases are described by the OPs with $L = 2$: $R_{jk} = \langle D_{jk}^L \rangle$. Consider the molecules that possess symmetry $C_{2v}$ or $D_{2h}$. The Schönflies symbol $C_{2v}$ is assigned to the point group with symmetry operations of identity, rotation around two-fold symmetry axis $C_2$, and two planes of mirror symmetry containing $C_2$ axis. The symbol $D_{2h}$ refers to the point group in which besides the symmetries above, there are two more $C_2$ rotation axes, inversion, and the planes of mirror symmetry perpendicular to $C_2$ axes. For these molecules, we introduce the molecular frame $\mathbb{M}$ with the axes $\hat{\mathbf{m}}_i$ parallel and perpendicular to the symmetry axis and symmetry plane. The nematic phase formed by these molecules features four independent OPs: two uniaxial OPs, denoted $R_{00}$, $R_{02} = R_{0-2}$, and two biaxial OPs, denoted $R_{20} = R_{-20}$, $R_{22} = R_{\pm 2 \pm 2}$, in the laboratory frame $\mathbb{L} = Oxyz$ defined by the directors [27-29], with $\hat{\mathbf{N}}_0 = (0,0,1)$. The OPs $R_{00}$ and $R_{20}$ describe, respectively, the uniaxial and biaxial orientational order of the long molecular axes $\hat{\mathbf{m}}_3$ and determine the diagonal form

$\left\{ -\left(R_{00} - \sqrt{6}R_{20}\right)/3, -\left(R_{00} + \sqrt{6}R_{20}\right)/3, 2R_{00}/3 \right\}$ of the traceless tensor OP $\mathbf{Q} = \langle \hat{\mathbf{m}}_3 \otimes \hat{\mathbf{m}}_3 \rangle - \mathbf{I}/3$ [1, 14] in the laboratory frame along the directors. The uniaxial OP $R_{00}$ is nothing else but the standard nematic OP $S$, $R_{00} = S$. The OPs $R_{02}$ and $R_{22}$ describe, respectively, the uniaxial and biaxial orderings of the short axes $\hat{\mathbf{m}}_{1,2}$ and are equivalent to the tensor $\mathbf{B} = \langle \hat{\mathbf{m}}_1 \otimes \hat{\mathbf{m}}_1 - \hat{\mathbf{m}}_2 \otimes \hat{\mathbf{m}}_2 \rangle$ [31], which has the diagonal form $\left\{ -\left(\sqrt{2}R_{02} - 2\sqrt{3}R_{22}\right)/3, -\left(\sqrt{2}R_{02} + 2\sqrt{3}R_{22}\right)/3, 2\sqrt{2}R_{02}/3 \right\}$ in the laboratory frame along the directors. Without the electric field, the NLC under consideration is uniaxial with the equilibrium uniaxial OPs $R_{00}^{(0)}$ and $R_{02}^{(0)}$, while the biaxial OPs are zero, $R_{20}^{(0)} = R_{22}^{(0)} = 0$. The electric field $\mathbf{E}$ changes the OPs $\delta R_{jk} = R_{jk} - R_{jk}^{(0)}$ through $\delta \boldsymbol{\varepsilon}^{(m)}$. When $\delta R_{jk}$ is small and the field is applied along one of the laboratory axes, the diagonal elements $\{\delta \varepsilon_x, \delta \varepsilon_y, \delta \varepsilon_z\}$ of the dielectric tensor $\delta \boldsymbol{\varepsilon}^{(m)}$ are

$$\delta \varepsilon_i = \sum_{j,k=0,2} \varepsilon_{i,jk} \delta R_{jk}, \quad i = x, y, z, \tag{3}$$

where $\varepsilon_{i,jk} = \partial \varepsilon_i / \partial(\delta R_{jk})\big|_{\varepsilon = \varepsilon^{(0)}}$. Rotation of $\mathbb{L}$ by $\pi/2$ around $Oz$ changes the sign of the biaxial OPs $\delta R_{2k}$ but does not affect the uniaxial OPs $\delta R_{0k}$. This results in the following properties: (a) $\varepsilon_{z,2k} = 0$ and, therefore, $\delta \varepsilon_z$ contains only the uniaxial OPs $\delta R_{0k}$, (b) the relation $\varepsilon_{y,jk} = (-1)^{j/2} \varepsilon_{x,jk}$ stands, (c) the quadratic expansion of microscopic $f_m$ near the zero-field equilibrium value $f_m^{(0)}$ with $R_{jk} = R_{jk}^{(0)}$ does not contain cross-terms of the uniaxial and biaxial OPs:

$$f_m = f_m^{(0)} + \frac{1}{2} \sum_{j,k,k'} M_{jk,jk'} \delta R_{jk} \delta R_{jk'}, \tag{4}$$

where $M_{jk,jk'} = \left( \partial^2 f_m / \partial R_{jk} \partial R_{jk'} \right)_{R_{jk} = R_{jk}^{(0)}}$ are the Taylor coefficients that can be determined from the Landau expansion of the free energy for uniaxial and biaxial nematics [32], indices $j$, $k$, and $k'$ run through two values 0 and 2. Because we consider processes with characteristic times less than a microsecond, the heat transfer is negligible [33], and, therefore, $M_{jk,jk'}$ corresponds to the expansion under adiabatic conditions.

We model the dynamics of the OPs $\delta R_{jk}$ using the standard Landau-Khalatnikov approach [34]:

$$\gamma_{jk}\frac{d(\delta R_{jk})}{dt} = -\frac{\partial(f_d + f_m)}{\partial(\delta R_{jk})} = G_{jk}E^2(t) - \sum_{k'} M_{jk,jk'}\delta R_{jk'}, \tag{5}$$

where $G_{jk} = \frac{\varepsilon_0}{2}\sum_i \varepsilon_{i,jk} e_i^2$, $\hat{\mathbf{e}}$ is the direction of the applied electric field $\mathbf{E}(t)$, and $\gamma_{jk}$ is the rotational viscosity for the OP $\delta R_{jk}$. We neglect the effects of the director reorientation and associated flows on the OPs, discussed in [35, 36], because we consider the geometries when the applied electric field stabilizes the director $\hat{\mathbf{N}}_0$. Four equations (5) are two independent pairs of linear inhomogeneous ordinary differential equations with constant coefficients for the uniaxial $\delta R_{0k}$ and biaxial $\delta R_{2k}$ OPs, and could be written in a vector form:

$$\boldsymbol{\xi}_{(j)}\frac{d}{dt}\mathbf{R}^{(j)} = \boldsymbol{\xi}_{(j)}^{-1}\mathbf{G}^{(j)}E^2(t) - \bar{\mathbf{M}}^{(j)}\boldsymbol{\xi}_{(j)}\mathbf{R}^{(j)}, \tag{6}$$

where $\mathbf{R}^{(j)} = \begin{pmatrix} \delta R_{j0} \\ \delta R_{j2} \end{pmatrix}$, $\mathbf{G}^{(j)} = \begin{pmatrix} G_{j0} \\ G_{j2} \end{pmatrix}$, $\boldsymbol{\xi}_{(j)} = \begin{pmatrix} \gamma_{j0}^{1/2} & 0 \\ 0 & \gamma_{j2}^{1/2} \end{pmatrix}$, and $\bar{\mathbf{M}}^{(j)}$ is the $2\times 2$ symmetric matrix with elements $\bar{M}_{kk'}^{(j)} = \gamma_{jk}^{-1/2} M_{jk,jk'} \gamma_{jk'}^{-1/2}$. Solution of Eq. (6) $\mathbf{R}^{(j)}(t)$ can be expressed through the vector of decoupled relaxation modes $\mathbf{r}^{(j)}(t) = \begin{pmatrix} r_0^{(j)}(t) \\ r_2^{(j)}(t) \end{pmatrix}$:

$$\mathbf{R}^{(j)}(t) = \boldsymbol{\xi}_{(j)}^{-1}\mathbf{V}^{(j)}\mathbf{r}^{(j)}(t), \tag{7}$$

where $\mathbf{V}^{(j)}$ is the matrix of eigenvectors of $\bar{\mathbf{M}}^{(j)}$ that obeys the equation $\bar{\mathbf{M}}^{(j)}\mathbf{V}^{(j)} = \mathbf{V}^{(j)}\boldsymbol{\Lambda}^{(j)}$; here $\boldsymbol{\Lambda}^{(j)} = \begin{pmatrix} \lambda_0^{(j)} & 0 \\ 0 & \lambda_2^{(j)} \end{pmatrix}$ is a diagonal matrix of the eigenvalues $\lambda_{0,2}^{(j)}$. Since $\bar{\mathbf{M}}^{(j)}$ is a symmetric positively defined matrix, $\mathbf{V}^{(j)} = \begin{pmatrix} \cos\phi_j & -\sin\phi_j \\ \sin\phi_j & \cos\phi_j \end{pmatrix}$ is an orthogonal matrix and is determined by the eigenvector angle $\phi_j$, which satisfies the equation

$$\tan 2\phi_j = 2\bar{M}_{02}^{(j)}/\left(\bar{M}_{00}^{(j)} - \bar{M}_{22}^{(j)}\right). \tag{8}$$

It is also convenient to use $\phi_j$ in expression for $\lambda_{0,2}^{(j)} = \frac{1}{2}\left[\bar{M}_{00}^{(j)} + \bar{M}_{22}^{(j)} \pm \left(\bar{M}_{00}^{(j)} - \bar{M}_{22}^{(j)}\right)/\cos 2\phi_j\right]$, because selection $|\phi_j| < \pi/4$ as the range of solutions of Eq.(8) ensures that the dynamics of the uniaxial OPs $\delta R_{jk}$ are mainly controlled by $r_k^{(j)}(t)$ with the corresponding relaxation time $\tau_k^{(j)} = 1/\lambda_k^{(j)}$:

$$r_k^{(j)}(t) = g_k^{(j)} \int_0^t E^2(t')\exp\left[(t'-t)/\tau_k^{(j)}\right] dt', \tag{9}$$

where $g_k^{(j)}$ are the components of the vector $\mathbf{g}^{(j)} = \left(\mathbf{V}^{(j)}\right)^{-1} \boldsymbol{\xi}_{(j)}^{-1} \mathbf{G}^{(j)}$.

To describe the optic manifestation of the NEMOP effect, we use the OPs-related deviation $\delta\tilde{\boldsymbol{\varepsilon}}^{(m)}$ of the dielectric tensor at optical frequency (optic tensor) from its zero-field value $\tilde{\boldsymbol{\varepsilon}}^{(0)}$. Here and in what follows, tildes represent a reference to the material parameters at the optical frequencies. In the laboratory frame $Oxyz$ along the directors, the tensor $\delta\tilde{\boldsymbol{\varepsilon}}^{(m)}$ has the diagonal form $\{\delta\tilde{\varepsilon}_x, \delta\tilde{\varepsilon}_y, \delta\tilde{\varepsilon}_z\}$ and can be split into an isotropic $\delta\tilde{\varepsilon}_{iso}$, uniaxial $\delta\tilde{\varepsilon}_u$, and biaxial $\delta\tilde{\varepsilon}_b$ contributions

$$\begin{aligned}
\delta\tilde{\varepsilon}_x &= \delta\tilde{\varepsilon}_{iso} - \frac{1}{3}\delta\tilde{\varepsilon}_u + \frac{1}{2}\delta\tilde{\varepsilon}_b, \\
\delta\tilde{\varepsilon}_y &= \delta\tilde{\varepsilon}_{iso} - \frac{1}{3}\delta\tilde{\varepsilon}_u - \frac{1}{2}\delta\tilde{\varepsilon}_b, \\
\delta\tilde{\varepsilon}_z &= \delta\tilde{\varepsilon}_{iso} + \frac{2}{3}\delta\tilde{\varepsilon}_u.
\end{aligned} \tag{10}$$

Since $\delta\boldsymbol{\varepsilon}^{(m)}$ and $\delta\tilde{\boldsymbol{\varepsilon}}^{(m)}$ are the same tensor at different frequencies, the deviations $\delta\tilde{\varepsilon}_i = \sum_{j,k=0,2} \tilde{\varepsilon}_{i,jk} \delta R_{jk}$ should be also linear in $\delta R_{jk}$, where $\tilde{\varepsilon}_{i,jk} = \partial\tilde{\varepsilon}_i/\partial\left(\delta R_{jk}\right)_{\tilde{\boldsymbol{\varepsilon}}=\tilde{\boldsymbol{\varepsilon}}^{(0)}}$ have the same symmetry properties as $\varepsilon_{i,jk}$. Then, the dynamics of $\delta\tilde{\varepsilon}_{iso}$ and $\delta\tilde{\varepsilon}_u$ are controlled by the uniaxial OPs $\delta R_{0k}$ and, therefore, by the vector of uniaxial modes $\mathbf{r}^{(0)}(t)$, whereas $\delta\tilde{\varepsilon}_b$ is controlled by the biaxial OPs $\delta R_{2k}$ and by $\mathbf{r}^{(2)}(t)$:

$$\delta\tilde{\varepsilon}_{iso}(t) = \tilde{\mathbf{h}}^{(iso)}\boldsymbol{\xi}_{(0)}^{-1}\mathbf{V}^{(0)}\mathbf{r}^{(0)}(t),$$
$$\delta\tilde{\varepsilon}_{u}(t) = \tilde{\mathbf{h}}^{(u)}\boldsymbol{\xi}_{(0)}^{-1}\mathbf{V}^{(0)}\mathbf{r}^{(0)}(t), \quad (11)$$
$$\delta\tilde{\varepsilon}_{b}(t) = \tilde{\mathbf{h}}^{(b)}\boldsymbol{\xi}_{(2)}^{-1}\mathbf{V}^{(2)}\mathbf{r}^{(2)}(t),$$

where $\tilde{\mathbf{h}}^{(iso)}$, $\tilde{\mathbf{h}}^{(u)}$ and $\tilde{\mathbf{h}}^{(b)}$ are vectors with components, respectively, $\tilde{h}_{k}^{(iso)} = \frac{1}{3}(\tilde{\varepsilon}_{x,0k} + \tilde{\varepsilon}_{y,0k} + \tilde{\varepsilon}_{z,0k})$, $\tilde{h}_{k}^{(u)} = \tilde{\varepsilon}_{z,0k} - (\tilde{\varepsilon}_{x,0k} + \tilde{\varepsilon}_{y,0k})/2$, and $\tilde{h}_{k}^{(b)} = \tilde{\varepsilon}_{x,2k} - \tilde{\varepsilon}_{y,2k}$.

The dynamics of the NEMOP effect is described by two uniaxial and two biaxial relaxation modes, Eqs. (9) and (11). When **E** is perpendicular to the $Oz$ axis (chosen parallel to the director) and $\Delta\varepsilon < 0$, all four modes should contribute to the optic response. However, as we will show below our experimental data for dielectrically negative material CCN-47 are fitted well by the simplified version of the model with one uniaxial mode and one biaxial mode. We explain this fact by the assumption that the NEMOP effect is controlled by the following two modes: (i) $r_0^{(0)}(t)$, associated mainly with the uniaxial OP $R_{00} = S$ of the long molecular axes, and (ii) $r_2^{(2)}(t)$, associated mainly with the biaxial OP $R_{22}$ of the short molecular axes. These two OPs are predicted to be dominant in the spontaneous (field-free) uniaxial and biaxial NLC [32, 37]. The same OPs are expected to play the major role in NEMOP experiments, since $\delta R_{00}$ causes strong changes in optic anisotropy (large $\tilde{h}_0^{(u)}$), and $\delta R_{22}$ is strongly affected by the interactions between the transverse molecular dipoles and the electric field (large $G_{22}$). In this two-mode assumption, the isotropic $\delta\tilde{\varepsilon}_{iso}$, uniaxial $\delta\tilde{\varepsilon}_{u}$, and biaxial $\delta\tilde{\varepsilon}_{b}$ contributions, Eq. (11), are simplified:

$$\delta\tilde{\varepsilon}_{\bar{j}}(t) = \frac{\alpha_{\bar{j}}}{\tau_{\bar{j}}}\int_0^t E^2(t')\exp\left[(t'-t)/\tau_{\bar{j}}\right]dt'. \quad (12)$$

where $\bar{j}$ reads $iso$, $u$, or $b$ depending on the nature of contribution, $\tau_u = \tau_{iso} = \tau_0^{(u)} \approx \gamma_{00}/M_{00,00}$ and $\tau_b = \tau_2^{(b)} \approx \gamma_{22}/M_{22,22}$ are the uniaxial and biaxial relaxation times, $\alpha_u \approx \tilde{h}_0^{(u)}G_{00}/M_{00,00}$ and $\alpha_b \approx \tilde{h}_2^{(b)}G_{22}/M_{22,22}$ are the effective uniaxial and biaxial susceptibilities, respectively. One can expect that $\tau_u$, determined by reorientation of the long axes, is substantially larger than $\tau_b$, determined by rotation of the short axes, because the former process is associated with the larger moment of inertia and requires stronger readjustment of the neighboring molecules. For the electric field parallel to the $Ox$ axis, $\hat{\mathbf{e}} = (1,0,0)$, one can estimate

$$\alpha_u \approx \varepsilon_0 \varepsilon_{x,00} \tilde{\varepsilon}_{x,00} / 2M_{00,00},$$
$$\alpha_b \approx \varepsilon_0 \varepsilon_{x,22} \tilde{\varepsilon}_{x,22} / M_{22,22}.$$
(13)

The uniaxial $\delta \tilde{\varepsilon}_u$ and biaxial $\delta \tilde{\varepsilon}_b$ terms provide the main contributions to NEMOP. The dynamics of the isotropic term $\delta \tilde{\varepsilon}_{iso}$ is similar to that of $\delta \tilde{\varepsilon}_u$, but its contribution is relatively small: $\delta \tilde{\varepsilon}_{iso} = 0$ under the assumption that $\tilde{\varepsilon}$ is an orientational average of the molecular polarizability tensor, because $\text{Tr}\tilde{\varepsilon} = \sum_i \tilde{\varepsilon}_i = const$ in this case [32], and the only non-zero contribution to $\delta \tilde{\varepsilon}_{iso}$ stems from the dipole-dipole resonance and dispersion intermolecular interactions [38]. Moreover, $\delta \tilde{\varepsilon}_{iso}$ does not contribute to the response caused by changes of birefringence.

## B. Dynamics of director fluctuations in electric field

Besides the NEMOP effect, the electric field provides an additional electro-optic response, which is of macroscopic nature. In NLCs with a negative dielectric anisotropy, the electric field $\mathbf{E} = (E,0,0)$ does not reorient the average $\hat{\mathbf{N}}_0 = (0,0,1)$ but modifies the director fluctuations $\delta \mathbf{N} = \hat{\mathbf{N}} - \hat{\mathbf{N}}_0$. We analyze this effect using the macroscopic part of free energy $\overline{F} = \int_V (f_e + \overline{f}_d) dV$, where $V = d \times L_y \times L_z$ is the active volume of the cell, covered by the electrodes of the area $L_y \times L_z$, and $d$ is the thickness of the NLC layer. The elastic energy density $f_e$ is

$$f_e = \frac{1}{2} \left[ K_1 (\text{div}\,\hat{\mathbf{N}})^2 + K_2 (\hat{\mathbf{N}} \cdot \text{curl}\,\hat{\mathbf{N}})^2 + K_3 (\hat{\mathbf{N}} \times \text{curl}\,\hat{\mathbf{N}})^2 \right],$$
(14)

where $K_1$, $K_2$, and $K_3$ are the Frank elasticity constants for splay, twist, and bend respectively. The dielectric energy density associated with the director distortions $\overline{f}_d = -\frac{1}{2}\varepsilon_0 \mathbf{E}\,\overline{\boldsymbol{\varepsilon}}\,\mathbf{E}$ is determined by the corresponding part of the dielectric tensor $\bar{\boldsymbol{\varepsilon}} = \boldsymbol{\varepsilon}^{(0)}\left(R_{jk}^{(0)}, \hat{\mathbf{N}}_0\right) + \delta\boldsymbol{\varepsilon}^{(fl)}\left(R_{jk}^{(0)}, \hat{\mathbf{N}}\right) = \varepsilon_\perp^{(0)}\mathbf{I} + \left(\varepsilon_\parallel^{(0)} - \varepsilon_\perp^{(0)}\right)\hat{\mathbf{N}} \otimes \hat{\mathbf{N}}$, where $\mathbf{I}$ is the unit tensor, $\varepsilon_\perp^{(0)}$ and $\varepsilon_\parallel^{(0)}$ are the dielectric constants, perpendicular and parallel to $\hat{\mathbf{N}}_0$, and $\otimes$ denotes the outer product.

We assume that the director fluctuations $\delta\mathbf{N} = \left(N_x(\mathbf{r}), N_y(\mathbf{r}), 0\right)$ are small, periodic in the $Oyz$ area of $V$, and obey the strong anchoring boundary conditions at the substrates. Thus we expand $\delta\mathbf{N}$ in Fourier series, similar to [39]:

$$\delta\mathbf{N}(\mathbf{r}) = \sum_{\mathbf{q}} \mathbf{N}(\mathbf{q}) \sin(q_x x) \exp\left[i(q_y y + q_z z)\right], \qquad (15)$$

where $\mathbf{q} = (q_x, q_y, q_z) = \left(\dfrac{\pi}{d}k, \dfrac{2\pi}{L_y}l, \dfrac{2\pi}{L_z}m\right)$ are discrete wavevectors with $k > 0$, $l$, and $m$ being integers.

Using Eq. (15) and integrating over $V$, we obtain $\bar{F}$ associated with the director fluctuations in the Gaussian approximation,

$$\bar{F} = \frac{V}{4}\sum_{\mathbf{q}}\left[\left(K_1 q_x^2 + K_2 q_y^2 + K_3 q_z^2 - \varepsilon_0 \Delta\varepsilon^{(0)} E_x^2\right)\left|N_x^2(\mathbf{q})\right| + \left(K_1 q_y^2 + K_2 q_x^2 + K_3 q_z^2\right)\left|N_y^2(\mathbf{q})\right|\right] + \\ + 2\pi i L_z (K_1 - K_2) \sum_{\mathbf{q},\mathbf{q}'}\left[N_x(\mathbf{q})N_y^*(\mathbf{q}') + N_x^*(\mathbf{q}')N_y(\mathbf{q})\right]\frac{klk'}{(k^2 - k'^2)}, \qquad (16)$$

where the latter sum contains the cross-terms of $N_x(\mathbf{q})$ and $N_y(\mathbf{q}')$, with $(k-k')$ being an odd number, $l = l'$, and $m = m'$.

To describe the dynamics of fluctuations, we start with the Langevin equation by including the random force $\zeta_\alpha(t,\mathbf{q})$ in the viscous relaxation equation for $N_\alpha(t,\mathbf{q})$, $\alpha = x, y$ [1, 16, 25], and use the splay-twist one-constant approximation $K_1 = K_2 = \bar{K}$, which diagonalizes the free energy, Eq. (16), with respect to $N_x(\mathbf{q})$ and $N_y(\mathbf{q})$,

$$\eta_\alpha(\hat{\mathbf{q}})\frac{dN_\alpha(t,\mathbf{q})}{dt} = -f_\alpha(t,\mathbf{q})N_\alpha(t,\mathbf{q}) + \zeta_\alpha(t,\mathbf{q}), \qquad (17)$$

where $f_y(t,\mathbf{q}) = f_K(\mathbf{q}) = \bar{K}(q_x^2 + q_y^2) + K_3 q_z^2$, $f_x(t) = f_K(\mathbf{q}) + f_E(t,\mathbf{q})$, $f_E(t) = \varepsilon_0|\Delta\varepsilon|E^2(t)$, and $\eta_\alpha(\hat{\mathbf{q}}) = \gamma_1 - \Delta\eta_\alpha(\hat{\mathbf{q}})$ is the effective director viscosity; here $\gamma_1$ is the director rotational viscosity and $\Delta\eta_\alpha(\hat{\mathbf{q}})$ is the backflow effect's correction, which, in the hydrodynamic limit of small $\mathbf{q}$, depends on $\hat{\mathbf{q}} = \mathbf{q}/|\mathbf{q}|$ [1, 16, 25]. The random force $\zeta_\alpha(t,\mathbf{q})$ has the standard 'white noise' properties with the noise strength $Z_\alpha(t,\mathbf{q})$

$$\langle \zeta_\alpha(t,\mathbf{q}) \rangle = 0, \quad \langle \zeta_\alpha(t,\mathbf{q}) \zeta_{\alpha'}^*(t',\mathbf{q}') \rangle = Z_\alpha(t,\mathbf{q}) \delta(t-t') \delta_{\mathbf{qq}'} \delta_{\alpha\alpha'}, \tag{18}$$

where the brackets $\langle ... \rangle$ denote an ensemble average. The solution of Eq.(17)

$$N_\alpha(t,\mathbf{q}) = e^{-S_\alpha(t,\mathbf{q})} \left[ N_\alpha(0,\mathbf{q}) + \eta_\alpha^{-1}(\hat{\mathbf{q}}) \int_0^t e^{S_\alpha(t',\mathbf{q})} \zeta_\alpha(t',\mathbf{q}) dt' \right], \tag{19}$$

where $S_\alpha(t,\mathbf{q}) = \eta_\alpha^{-1} \int_0^t f_\alpha(t',\mathbf{q}) dt'$, allows us to derive the equation that controls the dynamics of ensemble averaged fluctuations $\langle |N_\alpha^2(t,\mathbf{q})| \rangle$

$$\tau_\alpha(t,\mathbf{q}) \frac{d\langle |N_\alpha^2(t,\mathbf{q})| \rangle}{dt} = \frac{Z_\alpha(t,\mathbf{q})}{2\eta_\alpha(\hat{\mathbf{q}}) f_\alpha(t,\mathbf{q})} - \langle |N_\alpha^2(t,\mathbf{q})| \rangle, \tag{20}$$

where $\tau_\alpha(t,\mathbf{q}) = \eta_\alpha(\hat{\mathbf{q}})/2f_\alpha(t,\mathbf{q})$ is the characteristic relaxation time. For the stationary electric field $E$, the averaged fluctuations $\langle |N_\alpha^2(t,\mathbf{q})| \rangle$ can be calculated using the Equipartition Theorem and the free energy, Eq. (16), $\langle |N_\alpha^2(t,\mathbf{q})| \rangle_E = \frac{2k_B T}{V f_\alpha(t,q)}$, thus $Z_\alpha(t,\mathbf{q}) = \frac{2k_B T}{V} \eta_\alpha(\hat{\mathbf{q}})$.

The fluctuations along the $y$ axis are not affected by the applied field, $\langle |N_y^2(t,\mathbf{q})| \rangle = \langle |N_y^2(0,\mathbf{q})| \rangle$, and only the dynamics of $\langle |N_x^2(t,\mathbf{q})| \rangle$ affects the optic response. Introducing the field-induced quenching of fluctuations $\mathfrak{N}(t,\mathbf{q}) = \langle |N_x^2(0,\mathbf{q})| \rangle - \langle |N_x^2(t,\mathbf{q})| \rangle$, which satisfies the initial condition $\mathfrak{N}(0,\mathbf{q}) = 0$, we obtain the solution of Eq. (20) as

$$\mathfrak{N}(t,\mathbf{q}) = \frac{4k_B T}{V\eta_x(\hat{\mathbf{q}}) f_K(\mathbf{q})} \int_0^t f_E(t') \exp\left[ -\int_{t'}^t \frac{dt''}{\tau_x(\mathbf{q})} \right] dt'. \tag{21}$$

For a strong applied field, the electro-optic response is caused by the quenching of director fluctuations with a broad range of $\mathbf{q}$. Thus, we neglect the hydrodynamic effects and use an approximation of the constant effective rotational viscosity $\gamma_{eff}$ for the director fluctuations in the entire range of $\mathbf{q}$. In this case the solution, Eq. (21), is simplified to

$$\mathfrak{N}(t,\mathbf{q}) = \frac{2k_B T}{V\gamma_{eff} f_K(\mathbf{q})} e^{-S(t)} \int_0^t f_E(t') \exp\left[-\frac{f_K(\mathbf{q})(t-t')}{\gamma_{eff}}\right] e^{S(t')} dt', \tag{22}$$

where $S(t) = \dfrac{1}{\gamma_{eff}} \int_0^t f_E(t') dt'$ and $\gamma_{eff} \approx \gamma_1/2$.

Because the electric field affects only the director fluctuations along the $x$ axis, $\langle N_x^2(t,\mathbf{r})\rangle$, the associated modifications of the optic tensor are

$$\delta\tilde{\varepsilon}_z^{(fl)}(t,\mathbf{r}) = -\delta\tilde{\varepsilon}_x^{(fl)}(t,\mathbf{r}) = -\left(\langle N_x^2(t,\mathbf{r})\rangle - \langle N_x^2(0,\mathbf{r})\rangle\right)\left(n_e^2 - n_o^2\right), \tag{23}$$

where $n_o$ and $n_e$ are the ordinary and extraordinary refractive indices, respectively, measured in the field-free state, $E=0$.

In our experiments, we use a probing laser beam of half millimeter diameter and measure the phase retardation which is an integral along the cell thickness; thus, the fluctuations' contribution is determined by Eq. (23) averaged over the active volume of the cell

$$\delta\tilde{\varepsilon}_f(t) = V^{-1} \int_V \delta\tilde{\varepsilon}_z^{(fl)}(t,\mathbf{r}) d\mathbf{r} = \frac{\left(n_e^2 - n_o^2\right)}{2} \sum_{\mathbf{q}} \mathfrak{N}(t,\mathbf{q}). \tag{24}$$

The applied electric field affects the fluctuations, for which $q < q_c = \sqrt{\Delta\varepsilon\varepsilon_0 E^2/\bar{K}}$, as follows from the inequality $f_K(\mathbf{q}) < f_E$. For the strong electric field ($E \sim 10^8$ V/m), the number of these fluctuations is very large, as the maximum values of the integer indices are: $k_{max} > 10^3$ and $l_{max}, m_{max} > 10^5$. Thus, we neglect the discrete nature of $\mathbf{q}$ and transform the sum, Eq. (24), into an integral, where we stretch $q_z$, $\mathbf{q} \to \bar{\mathbf{q}} = \left(q_x, q_y, \sqrt{K_3/\bar{K}} q_z\right)$. This transformation makes the elastic term $f_K(\bar{\mathbf{q}}) = \bar{K}\bar{q}^2$ isotropic and, therefore, $\mathfrak{N}(t,\bar{q})$ also becomes isotropic:

$$\delta\tilde{\varepsilon}_f(t) = \left(n_e^2 - n_o^2\right) \frac{V\sqrt{\bar{K}}}{8\pi^3 \sqrt{K_3}} \int_{V_{\bar{\mathbf{q}}}} \mathfrak{N}(t,\bar{q}) d\bar{\mathbf{q}}, \tag{25}$$

where the integration volume $V_{\bar{\mathbf{q}}}$ is defined by conditions $\bar{q}_x \geq \pi/d$ and $\bar{q} < q_c = \pi/a_c$. Here the former condition stems from the strong anchoring at the substrates and $a_c$ is the characteristic distance that corresponds to the breakdown of continuum theory. Integrating (25) using (22), we obtain the contribution of the field-quenched director fluctuations to modification of the optic tensor:

$$\delta\tilde{\varepsilon}_f(t) = A\frac{e^{-S(t)}}{\sqrt{\gamma_{eff}}}\int_0^t \frac{f_E(t')e^{S(t')}}{\sqrt{t-t'}}\left\{\mathrm{erf}\sqrt{\frac{t-t'}{\tau_c}} - \mathrm{erf}\sqrt{\frac{t-t'}{\tau_d}} - \sqrt{\frac{t-t'}{\pi\tau_d}}\left[\mathrm{E}_1\left(\frac{t-t'}{\tau_d}\right) - \mathrm{E}_1\left(\frac{t-t'}{\tau_c}\right)\right]\right\}dt', \quad (26)$$

where $A = (n_e^2 - n_o^2)\dfrac{k_B T}{2\pi^{3/2}\bar{K}\sqrt{K_3}}$, $\tau_d = \dfrac{\gamma_{eff}d^2}{\bar{K}\pi^2} \approx \tau_{off}^F$, $\tau_c = \dfrac{\gamma_{eff}}{K_3 q_c^2}$, and $\mathrm{E}_1(t) = \int_t^\infty \dfrac{e^{-t'}}{t'}dt'$ $(t>0)$ is the exponential integral, see, e.g., chapter 5 of Ref.[40].

### C. Analysis and optimization of experimental geometries

We describe optical properties using the normalized wavevectors $\tilde{\mathbf{k}} = \dfrac{\lambda}{2\pi}\mathbf{k}$ of the optical modes, where $\lambda$ is the wavelength of a probing beam in vacuum. The tangential components $\tilde{k}_y$ and $\tilde{k}_z$ are preserved at interfaces between different layers: glass, ITO, polymer, nematic, etc., and are the same for all optical modes. The optical retardance between the two forward modes propagating through the field-induced (effectively biaxial) states of an NLC, $\Gamma = \Delta n_{eff} d$, is determined by the NLC thickness $d$ and the effective birefringence $\Delta n_{eff} = \tilde{k}_x^{(1)} - \tilde{k}_x^{(2)}$, where $\tilde{k}_x^{(1)}$ and $\tilde{k}_x^{(2)}$ are solutions of the Fresnel equation for two forward propagating modes, $\tilde{k}_x > 0$ in the biaxial medium:

$$\tilde{\varepsilon}_x \tilde{k}_x^4 - Q_2 \tilde{k}_x^2 + Q_0 = 0, \qquad (27)$$

where $Q_2 = \tilde{\varepsilon}_x(\tilde{\varepsilon}_y + \tilde{\varepsilon}_z) - \tilde{k}_y^2(\tilde{\varepsilon}_x + \tilde{\varepsilon}_y) - \tilde{k}_z^2(\tilde{\varepsilon}_x + \tilde{\varepsilon}_z)$ and $Q_0 = (\tilde{\varepsilon}_y\tilde{\varepsilon}_z - \tilde{\varepsilon}_y\tilde{k}_y^2 - \tilde{\varepsilon}_z\tilde{k}_z^2)(\tilde{\varepsilon}_x - \tilde{k}_y^2 - \tilde{k}_z^2)$.

In the field-free uniaxial state, modes 1 and 2 are the extraordinary $\tilde{k}_x^{(1)} = k_{xe} = \sqrt{n_e^2\left(1 - \dfrac{\tilde{k}_z^2}{n_o^2}\right) - \tilde{k}_y^2}$ and ordinary $\tilde{k}_x^{(2)} = k_{xo} = \sqrt{n_o^2 - \tilde{k}_y^2 - \tilde{k}_z^2}$ waves, respectively. An applied electric field causes a change of the effective birefringence $\delta n_{eff} = (\tilde{k}_x^{(1)} - k_{xe}) - (\tilde{k}_x^{(2)} - k_{xo})$, calculated from Eq. (27)

$$\delta n_{eff} = \frac{\delta\tilde{\varepsilon}_x \tilde{k}_{xe}\left(\tilde{k}_z^2 \tilde{k}_{xe}\tilde{k}_{xo} - \tilde{k}_y^2 n_o^2\right) + \delta\tilde{\varepsilon}_y \left(\tilde{k}_y^2 \tilde{k}_z^2 \tilde{k}_{xo} - \tilde{k}_{xe} n_o^2 \tilde{k}_{xo}^2\right) + \delta\tilde{\varepsilon}_z \left(n_o^2 - \tilde{k}_z^2\right)^2 \tilde{k}_{xo}}{2\tilde{k}_{xe}\tilde{k}_{xo} n_o^2 \left(n_o^2 - \tilde{k}_z^2\right)}. \quad (28)$$

The optic tensor modifications $\delta\tilde{\varepsilon}_x$, $\delta\tilde{\varepsilon}_y$, and $\delta\tilde{\varepsilon}_z$ contain the uniaxial $\delta\tilde{\varepsilon}_u$ and isotropic $\delta\tilde{\varepsilon}_{iso}$ contributions associated with the field-enhanced uniaxial order, the term stemmed from the field-induced biaxial order $\delta\tilde{\varepsilon}_b$, and the contribution $\delta\tilde{\varepsilon}_f$ caused by the quenching of director fluctuations along the $x$ axis. In real samples, there is also an additional 'pretilt' term, because the surface alignment direction at the bounding plates is practically never strictly parallel to the plate due to the small 'pretilt' angle $\beta$ induced by rubbing of the aligning layer. Nonzero $\beta$ implies that the zero-field director and the field are not strictly orthogonal, and that there is a nonzero dielectric torque on the director. The corresponding change in the effective birefringence is proportional to $\left(\bar{\beta}-\bar{\beta}_0\right)$, where $\bar{\beta}$ and $\bar{\beta}_0$ are the averaged angles between the director and the substrate plane with and without the applied electric field, respectively. One can show that $\bar{\beta}_0$ is the arithmetic mean of the pretilt angles at the top and bottom plates.

Using Eqs. (10) and (23) for the discussed contributions, we obtain from Eq. (28)

$$\delta n_{eff} = \sigma_{bu}\left(\delta\tilde{\varepsilon}_u + \frac{3}{2}\delta\tilde{\varepsilon}_b\right) + \sigma_{uf}\left(\delta\tilde{\varepsilon}_u + \frac{3}{2}\delta\tilde{\varepsilon}_f\right) + \sigma_\beta\left(\bar{\beta}-\bar{\beta}_0\right), \quad (29)$$

where $\sigma_{bu} = \frac{1}{6n_o^2\left(n_o^2 - \tilde{k}_z^2\right)}\left[\tilde{k}_z^2\left(\tilde{k}_{xe} - \frac{\tilde{k}_y^2}{\tilde{k}_{xe}}\right) + n_o^2\left(\tilde{k}_{xo} - \frac{\tilde{k}_y^2}{\tilde{k}_{xo}}\right)\right]$, $\sigma_{uf} = \frac{1}{3n_o^2}\left[\frac{n_o^2 - \tilde{k}_z^2}{\tilde{k}_{xe}} + \frac{\tilde{k}_y^2 n_o^2 - \tilde{k}_z^2 \tilde{k}_{xo}\tilde{k}_{xe}}{\tilde{k}_{xo}\left(n_o^2 - \tilde{k}_z^2\right)}\right]$,

and $\sigma_\beta = \frac{n_e^2 - n_o^2}{n_o^2}\tilde{k}_z$ are the weighting coefficients dependent on an experimental geometry. Note that $\delta\tilde{\varepsilon}_{iso}$ does not contribute to $\delta n_{eff}$ and therefore cannot be extracted from the phase retardance measurements. We also cannot completely separate $\delta\tilde{\varepsilon}_u$, $\delta\tilde{\varepsilon}_b$, and $\delta\tilde{\varepsilon}_f$ by staging three different experimental geometries, because these terms appear in Eq. (29) in two combinations. However, as we shall show below, there is a possibility to determine $\delta\tilde{\varepsilon}_u$, $\delta\tilde{\varepsilon}_b$, and $\delta\tilde{\varepsilon}_f$ independently utilizing their distinct dynamics.

We perform experiments for the following three geometries that provide the simplest interpretation:

(a) "Biaxial-uniaxial" (BU) geometry, in which the contribution of director fluctuations is eliminated, $\sigma_{uf} = 0$, and only the biaxial and uniaxial OPs contribute to the optic response.

(b) "Uniaxial-fluctuations" (UF) geometry: only the uniaxial OPs and director fluctuations contribute to the optic response, while the biaxial contribution does not, $\sigma_{bu} = 0$.

(c) "Normal" (N) geometry, with the perpendicular incidence of a probing beam, in which case all the three mechanisms (uniaxial, biaxial, and fluctuation-quenching) contribute to the measured signal, but the experimental setting and weighting coefficients in Eq. (29) are simple.

### 1. Biaxial-Uniaxial geometry

The simplest of the BU geometries, that satisfies the condition $\sigma_{uf} = 0$, is the one in which the incidence plane of a probing beam contains the director, $\tilde{k}_y = 0$, and the incidence angle obeys the condition $\tilde{k}_z = \frac{n_o^2}{\sqrt{n_e^2 + n_o^2}}$, Fig. 1(a). The field-induced change $\delta n_{BU}$ for this BU geometry is

$$\delta n_{BU} = \frac{n_o/n_e + 1 + n_e/n_o}{6\sqrt{n_e^2 + n_o^2}} \left( \delta\tilde{\varepsilon}_u + \frac{3}{2}\delta\tilde{\varepsilon}_b \right) + \frac{n_e^2 - n_o^2}{\sqrt{n_e^2 + n_o^2}} \left( \bar{\beta} - \bar{\beta}_0 \right). \tag{30}$$

The last term is a potential contribution of the finite pretilt angle at the boundaries. Because of the finite pretilt, the applied field can realign the director,

$$\bar{\beta}(t_{on} \leq t \leq t_{off}) = \bar{\beta}_0 \exp\left( -\frac{t - t_{on}}{\tau_{on}^F} \right), \tag{31}$$

where $\bar{\beta}_0$ is the arithmetic mean of the pretilt angles at the top and bottom plates when there is no field. After the field is switched off, the director relaxes back to the initial state,

$$\bar{\beta}(t > t_{off}) = \bar{\beta}_0 - \left[ \bar{\beta}_0 - \bar{\beta}(t_{off}) \right] \exp\left( -\frac{t - t_{off}}{\tau_{off}^F} \right). \tag{32}$$

At a timescale (1-1000) ns of interest, Eq. (32) yields a practically constant value of $\bar{\beta}(t > t_{off})$.

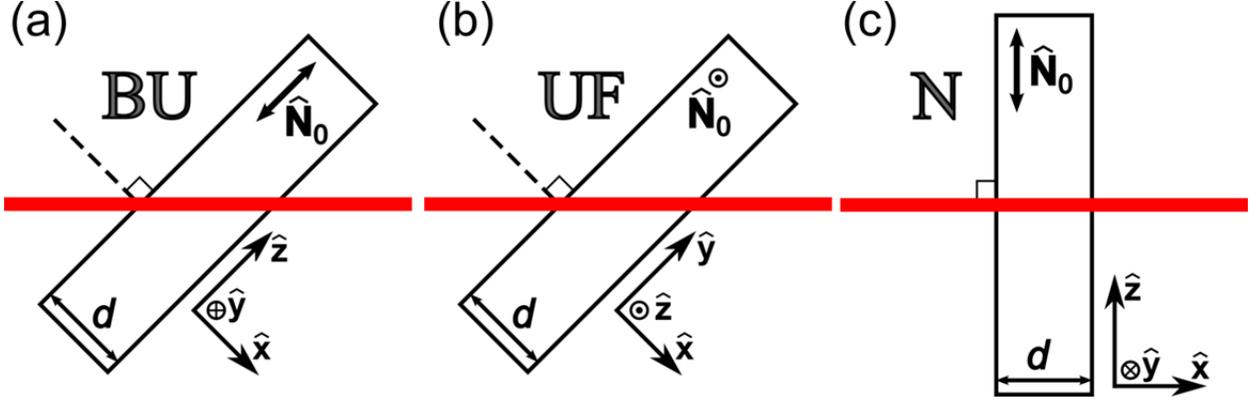

FIG. 1. (Color online) Three experimental schemes for testing an electro-optic response of a nematic cell with the laser beam (horizontal red line). (a) BU geometry probing biaxial and uniaxial contributions to the optic response. (b) UF geometry probing uniaxial and fluctuations quenching modifications. (c) N geometry, all three mechanisms contribute to the optic response.

### 2. Uniaxial-Fluctuative geometry

Among the UF geometries, determined by the condition $\sigma_{bu}=0$ in Eq. (29), we choose the one with the incidence plane of a probing beam perpendicular to the director, $\tilde{k}_z=0$, and the incidence angle obeying the condition $\tilde{k}_y = n_o/\sqrt{2}$, Fig. 1(b). The corresponding field-induced birefringence $\delta n_{UF}$ is

$$\delta n_{UF} = \frac{1}{3\sqrt{2}}\left(\frac{1}{n_o} + \frac{2}{\sqrt{2n_e^2 - n_o^2}}\right)\left(\delta\tilde{\varepsilon}_u + \frac{3}{2}\delta\tilde{\varepsilon}_f\right). \qquad (33)$$

If the refractive indices of NLC $n_e$ and $n_o$ are close to the refractive index of the glass substrate $n_g$, then the incident angles in BU and UF geometries are close to 45 degrees.

### 3. Normal geometry

In N geometry the probing light is perpendicular to the cell, $\tilde{k}_y = \tilde{k}_z = 0$, and Eq. (29) reduces to

$$\delta n_N = \frac{1}{6n_o}\left(\delta\tilde{\varepsilon}_u + \frac{3}{2}\delta\tilde{\varepsilon}_b\right) + \frac{1}{3n_e}\left(\delta\tilde{\varepsilon}_u + \frac{3}{2}\delta\tilde{\varepsilon}_f\right). \qquad (34)$$

## III. EXPERIMENTAL METHODS

We used commercially available NLC 4'-butyl-4-heptyl-bicyclohexyl-4-carbonitrile (CCN-47) (Nematel GmbH). The material parameters measured at $T=40°C$ are as following: dielectric constants $\varepsilon_\parallel = 3.9$, $\varepsilon_\perp = 9.0$, dielectric anisotropy $\Delta\varepsilon = -5.1$, all determined within the field frequency range 1-50 kHz; birefringence $\Delta n = 0.029$ at $\lambda = 633$ nm. The transverse dipole of CCN-47 molecules is large, $\mu_D = 12.3 \times 10^{-30}$ Cm $(3.7\,\text{Debye})$, as calculated using ChemOffice™ software. The structural formula of CCN-47 is shown in Fig. 2(a).

The cells were constructed from two parallel glass plates separated by spacers. The inner surfaces of these plates contain indium tin oxide (ITO) electrodes and unidirectionally rubbed polyimide layers PI-2555 (HD MicroSystems), which is separated by a gap $d$ in the range $(3.5-8.2)\,\mu m$. When a voltage pulse $U(t)$ is applied, an electric field $E(t)$ inside the liquid crystal is controlled by the $RC$-circuit, Fig. 2(b), formed by the resistance $R$ of the electrodes and the equivalent capacitance $C = C_{NLC}C_P/(C_{NLC} + C_P)$ created by the capacitances of the NLC $C_{NLC}$ and the polymer films $C_P$. Most of the experiments were performed with an NLC cell of the thickness $d = 4.2\,\mu m$ and the $RC$-time $\tau_{RC} = RC = 7$ ns. In order to reduce the $RC$-time, we used the electrodes of low resistivity $(10\,\Omega/\text{sq})$ and a small area, $A_e = 3 \times 3\,\text{mm}^2$, Fig. 2(c). The dielectric constant of the polyimide PI-2555 is $\varepsilon_P = 3.5$ [41]. The effective thickness for the capacitor formed by the two polymer films is $d_P = 0.2\,\mu m$. The rubbing directions at the plates are parallel to each other in order to minimize the effects of nonzero pretilt. The typical pretilt angle at the used substrates was about 0.7 degrees. To satisfy the conditions of the BU and UF geometries, Fig. 1, the NLC cell is sandwiched between two right angle glass prisms with the refractive index $n_g = 1.52$, which is close to $n_e = 1.50$ and $n_o = 1.47$ measured at $T = 40°C$ and $\lambda = 633$ nm. The temperature of the cells was controlled with accuracy 0.1°C by LTS350 hotstage (Linkam Scientific Instruments) and Linkam TMS94 controller.

The cells were tested with a He-Ne laser beam ($\lambda = 632.8$ nm), linearly polarized along the direction that makes an angle 45° with the incidence plane. The beam passes through the cell, the Soleil-Babinet compensator, and two crossed polarizers, Fig. 2(d). The transmitted light intensity was measured using a photodetector TIA-525 (Terahertz Technologies, response time < 1 ns).

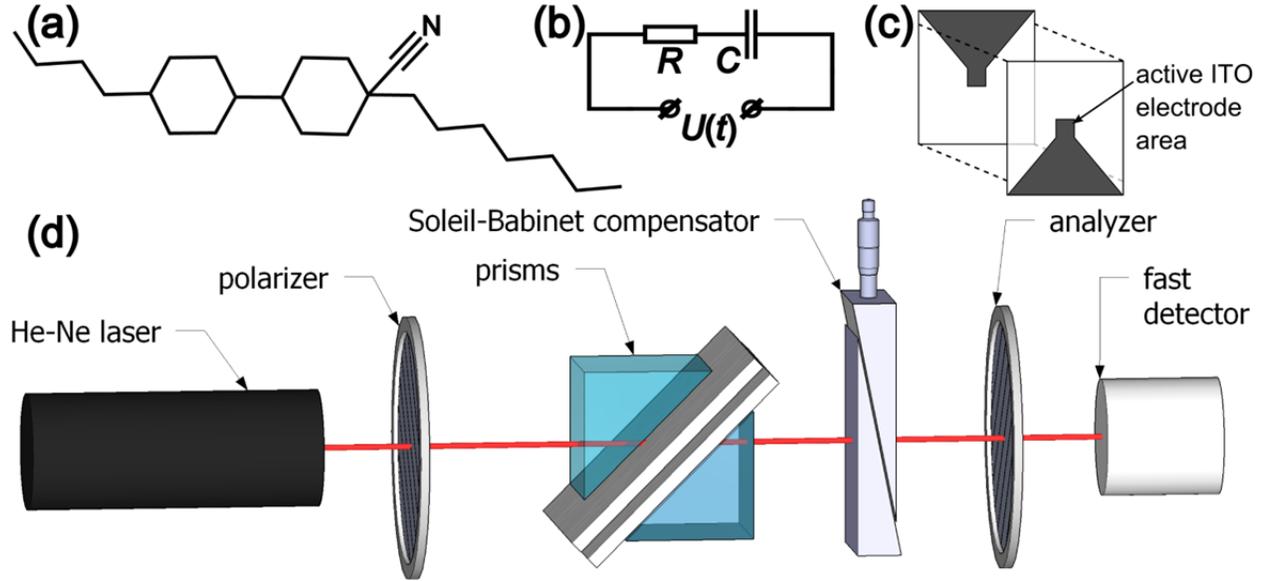

**FIG. 2.** (Color online) (a) Molecular structure of CCN-47. (b) Schematic *RC*-circuit. (c) Design of cell electrodes. (d) Electro-optic setup for geometries BU and UF.

The change in light intensity caused by the applied field can be presented as

$$I(t) = \left[I_{max}(t) - I_{min}(t)\right]\sin^2\left\{\frac{\pi\left[\delta n(t) + \Delta n_{eff}\right]d}{\lambda} + \frac{\phi_{SB}}{2}\right\} + I_{min}(t), \qquad (35)$$

where $\phi_{SB}$ is the variable phase retardance controlled by the Soleil-Babinet compensator, $I_{min}$ and $I_{max}$ are the minimum and maximum values of light intensity, respectively. The values of $I_{min}$ and $I_{max}$ are different from 0 and the ideal maximum because of parasitic effects such as light reflection at interfaces, light scattering, and absorption. These parasitic effects might be sensitive to the applied field, which is why both $I_{max}$ and $I_{min}$ are shown as time dependent in Eq. (35). The role of the variable Soleil-Babinet phase difference $\phi_{SB}$ is to eliminate the contribution of these parasitic effects from the effects affecting the birefringence, i.e., the OPs modifications and quenching of the director fluctuations, as explained below.

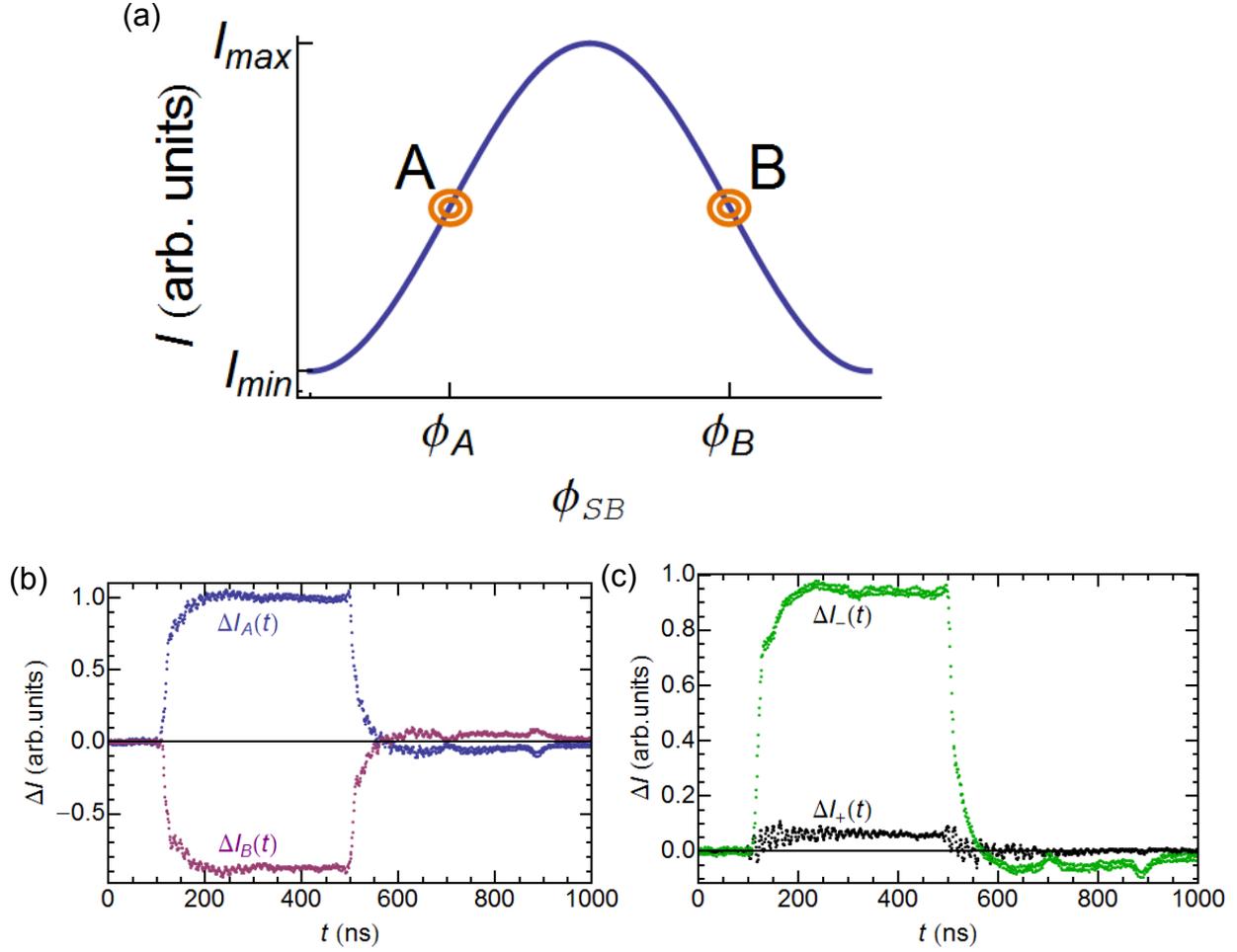

**FIG. 3.** (Color online) (a) Two settings of the Soleil-Babinet compensator, A and B, which correspond to the maximum sensitivity of light intensity to changes in optical retardance. The two settings also allow one to separate the field-induced retardance changes from parasitic effects. (b) The optic response to $U_0 = 626$ V pulse measured at $T = 43$°C, $d = 4.2$ μm for the two settings of the compensator, $\phi_{SB} = \phi_A$ and $\phi_{SB} = \phi_B$. (c) Half-difference $\Delta I_-(t)$, and half-sum $\Delta I_+(t)$ of the two optic response curves shown in Fig. 3(b).

The measurements are performed with two different values of the Soleil-Babinet phase retardation, $\phi_A = \frac{2\pi}{\lambda}\left(\frac{\lambda}{4} - \Delta n_{eff} d\right)$ and $\phi_B = \frac{2\pi}{\lambda}\left(\frac{3\lambda}{4} - \Delta n_{eff} d\right)$. At these values, the transmitted light intensity in the field-free state is $I(t=0) = \left[I_{max}(0) + I_{min}(0)\right]/2$, Fig. 3(a), which means that the sensitivity of light intensity to the changes of optical properties is maximized.

Furthermore, extraction of the useful contribution from the parasitic effects is achieved by evaluating the half-difference $\Delta I_-(t) = \frac{1}{2}[\Delta I_A(t) - \Delta I_B(t)] = \frac{\pi \delta n(t) d}{\lambda}[I_{max}(0) - I_{min}(0)]$ and the half-sum $\Delta I_+(t) = \frac{1}{2}[\Delta I_A(t) + \Delta I_B(t)] = \frac{1}{2}[\Delta I_{max}(t) + \Delta I_{min}(t)]$ of the optical measurements recorded for $\phi_A$ and $\phi_B$, Fig. 3(a) and 3(c). As seen in Fig. 3(c), the half-difference $\Delta I_-(t)$ signal is significantly larger than the half-sum $\Delta I_+(t)$ signal, which indicates the prevalence of the field-induced birefringence $\delta n(t)$ effect over the parasitic factors.

Voltage pulses of amplitude $U_0$ up to 1 kV, with nanoseconds' rise and fall fronts, were produced by a pulse generator HV 1000 (Direct Energy Inc). The profiles of voltage pulses $U(t)$ and optic responses $I(t)$ were experimentally determined with an oscilloscope Tektronix TDS 2014 (sampling rate 1GSample/s).

### IV.  OPTIC RESPONSE DYNAMICS AND EXPERIMENTAL DATA FITTING

Short voltage pulses of duration 394 ns applied to the NLC cell, Fig. 4(a), produce the optic responses shown in Figs. 4(b,c,d) for geometries BU, UF, and N, respectively.

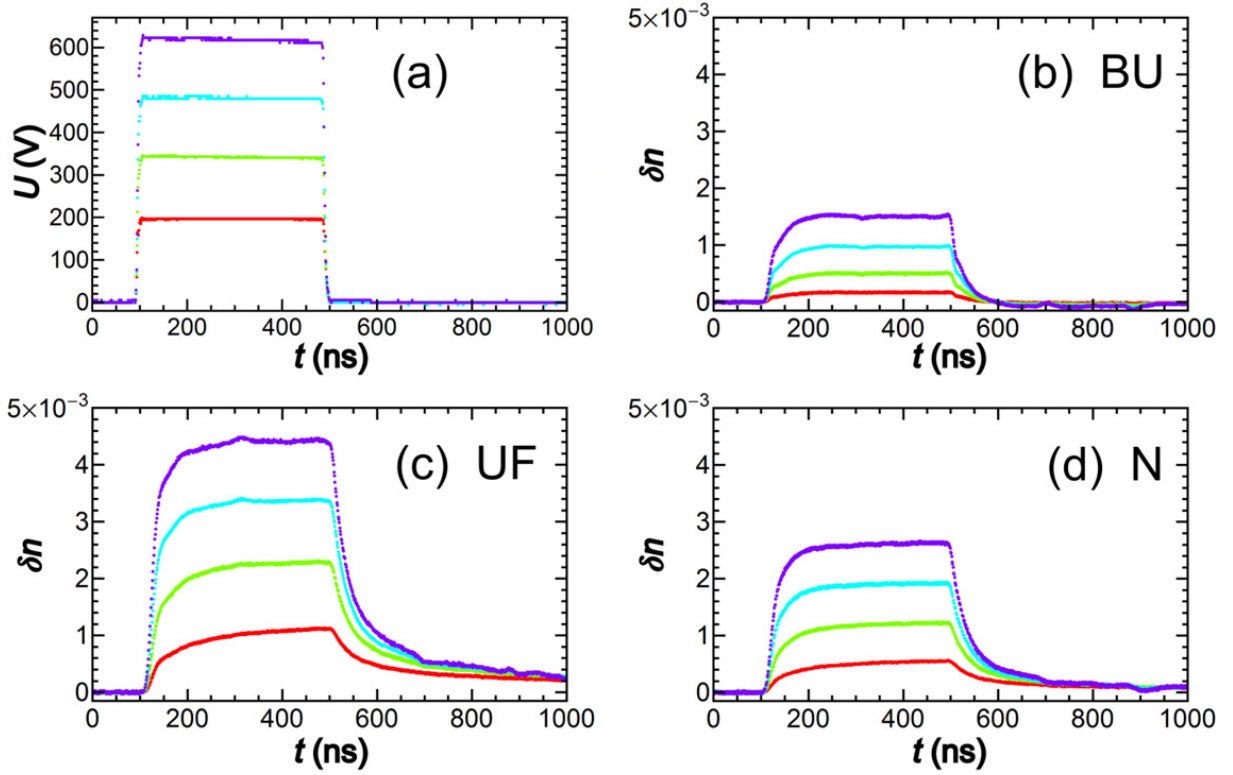

**FIG. 4.** (Color online) Dynamics of field-induced birefringence in geometries BU (b), UF (c), and N (d) in response to the applied voltage pulses (a); temperature $T = 49°C$. The curves in (a)-(d) from top to bottom correspond to voltage pulses with $U_0 = 626$ V, 484 V, 344 V, and 197 V, respectively.

In order to evaluate the dynamics of an optic response and to separate different contributions, one needs to know the profile of the voltage pulse. The latter can be presented as a sum of the exponential functions:

$$\begin{aligned} U(t < t_{on}) &= 0, \\ U(t_{on} \leq t \leq t_{off}) &= U_0 \left( e^{-(t-t_{on})/\tau_a} - e^{-(t-t_{on})/\tau_{on}} \right), \\ U(t > t_{off}) &= U(t_{off}) e^{-(t-t_{off})/\tau_{off}}, \end{aligned} \quad (36)$$

where $t_{on}$ and $t_{off}$ are the moments of time when the voltage is switched on and off, respectively; $U_0$ is the characteristic amplitude of the pulse applied to the electrodes of the cells, $\tau_{on}$ is the characteristic rise time of the front edge of the pulse, $\tau_{off}$ is the characteristic decay time of the rear edge of the pulse, and $\tau_a$ is the characteristic time of the slowly decaying amplitude of the pulse. The parameters $U_0$, $\tau_a$, $\tau_{on}$, and $\tau_{off}$ are obtained by fitting the experimental profile, Fig. 5(a). It is convenient to represent the voltage pulse as a sum of the exponential functions, because it allows us to solve the Kirchhoff equation for an *RC*-circuit with characteristic time $\tau_{RC}$, which is 7 ns for the cell of thickness 4.2 µm. Thus, the electric field inside the NLC $E(t<t_{on})=0$, $E^{ON}(t_{on} \leq t \leq t_{off})$, and $E^{OFF}(t>t_{off})$ is

$$E^{ON}(t_{on} \leq t \leq t_{off}) = E_0 \sum_i a_i e^{-\nu_i(t-t_{on})},$$

$$E^{OFF}(t>t_{off}) = E^{ON}(t_{off}) \sum_j b_j e^{-\mu_j(t-t_{off})},$$

(37)

where $E_0 = U_0 \varepsilon_P / (\varepsilon_\perp d_P + \varepsilon_P d)$. In our experiment for the switching-on dynamics, $t_{on} \leq t \leq t_{off}$, the summation index *i* runs through the values 1, 2, and 3; $a_i$ and $\nu_i$ are presented in Table I. And for the switching-off dynamics, $t > t_{off}$, the summation index *j* runs through the values 1 and 2; $b_j$ and $\mu_j$ are presented in Table II.

The exponential form representation of $E(t)$ streamlines the fitting procedure, because it allows one to evaluate Eq. (12) in an analytic form for the uniaxial $\delta\tilde{\varepsilon}_u(t)$ and biaxial $\delta\tilde{\varepsilon}_b(t)$ OPs dynamics as well as Eq. (26) for the quenching of director fluctuations $\delta\tilde{\varepsilon}_f(t)$.

**TABLE I**. Coefficients $a_i$ and $\nu_i$ for exponential expansion of $E^{ON}(t)$.

| $i$ | 1 | 2 | 3 |
|---|---|---|---|
| $a_i$ | $\dfrac{\tau_a}{\tau_a - \tau_{RC}}$ | $\dfrac{-\tau_{on}}{\tau_{on} - \tau_{RC}}$ | $\dfrac{-(\tau_a - \tau_{on}) \tau_{RC}}{(\tau_a - \tau_{RC})(\tau_{RC} - \tau_{on})}$ |
| $\nu_i$ | $1/\tau_a$ | $1/\tau_{on}$ | $1/\tau_{RC}$ |

**TABLE II.** Coefficients $b_j$ and $\mu_j$ for exponential expansion of $E^{OFF}(t)$.

| $j$ | 1 | 2 |
|---|---|---|
| $b_j$ | $1 - \dfrac{\varepsilon_P U(t_{off})}{(\varepsilon_\perp d_P + \varepsilon_P d) E^{ON}(t_{off})} \dfrac{\tau_{off}}{\tau_{off} - \tau_{RC}}$ | $\dfrac{\varepsilon_P U(t_{off})}{(\varepsilon_\perp d_P + \varepsilon_P d) E^{ON}(t_{off})} \dfrac{\tau_{off}}{\tau_{off} - \tau_{RC}}$ |
| $\mu_j$ | $1/\tau_{RC}$ | $1/\tau_{off}$ |

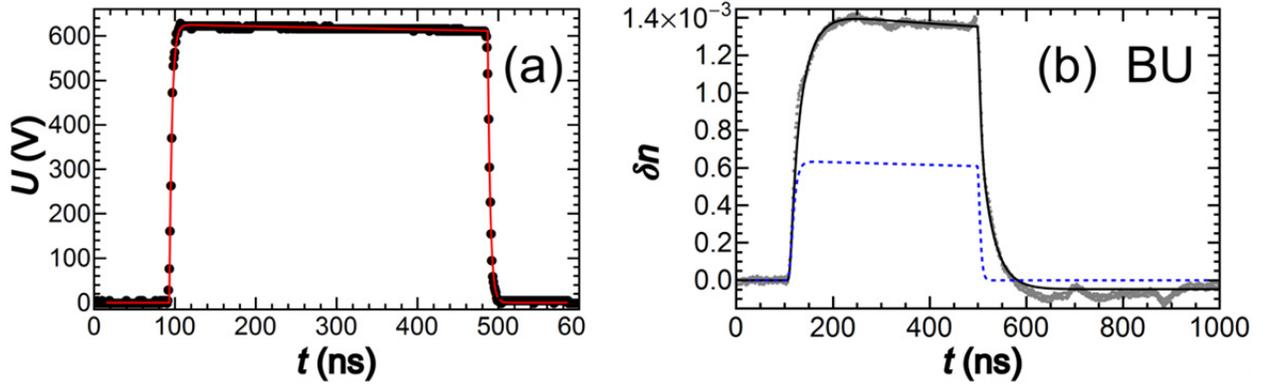

**FIG. 5**. (Color online) (a) Experimentally measured voltage profile fitted by Eq. (36) (solid red line) with $U_0 = 626$ V, $\tau_a = 18.5\ \mu s$, $\tau_{on} = 3.2$ ns, $\tau_{off} = 3.2$ ns, $t_{on} = 93$ ns, and $t_{off} = 487$ ns. (b) Optic response in BU geometry at $T = 46°C$ (gray dots) fitted with Eqs. (12), (30), (31), and (32) for one uniaxial and one biaxial mode, $\tau_b = 1.95$ ns, $\tau_u = 29$ ns, $\alpha_b = 5.4 \times 10^{-20}$ m$^2$/V$^2$, $\alpha_u = 8.9 \times 10^{-20}$ m$^2$/V$^2$, $\bar{\beta}_0 = 0.06°$, and $\tau_{on}^F = 85$ ns (solid black line). The blue dashed line is the biaxial contribution.

### A. Biaxial-Uniaxial geometry fitting

The typical response of CCN-47 to the applied voltage pulse of a duration of 394 ns, recalculated in terms of the field-induced birefringence change $\delta n$, is fitted according to Eq. (30), Fig. 5(b). The last term in Eq. (30) is the contribution due to the non-zero averaged pretilt angle $\bar{\beta}(t)$, which is described by Eqs. (31) and (32). We extract this contribution, using $\tau_{on}^F$ and considering that $\bar{\beta}(t) = \bar{\beta}(t_{off})$ is responsible for remaining a constant bias when in the range of 500-1000 ns, and Eq. (32) yields a practically constant value of $\bar{\beta} = 0.1°$. Two main contributions are the field-induced uniaxial $\delta\tilde{\varepsilon}_u(t)$ and biaxial $\delta\tilde{\varepsilon}_b(t)$ contributions of the NEMOP effect. Experimental data in the middle of the nematic phase fit well with the simplified model with two OPs, Eq. (12), and the fitting clearly reveals two processes with substantially different relaxation times: 'slow' in the range of tens of nanoseconds and 'fast' in the range of nanoseconds. We assign the 'slow' process, with relaxation time $\tau_u = 28\,\text{ns}$, to the uniaxial OP of long axes $\delta R_{00}$ and the 'fast' process, with $\tau_b = 1.95\,\text{ns}$, to the biaxial OP of short axes $\delta R_{22}$. This assignment is assigned by the experimental results for UF geometry, discussed in the next section. Although the experimental data should be generally discussed with four OPs, the data analysis shows that it suffices to use just two different OPs, and that the introduction of the third and fourth OP does not improve the fitting.

The experimental data, fitted with four parameters $\alpha_b$, $\alpha_u$, $\tau_b$, and $\tau_u$, clearly demonstrate that $\tau_b$ is the shortest timescale of the dynamic processes, being on the order of a few nanoseconds or even shorter. For all temperatures, the fitted values of $\tau_b$ are always shorter than 2.4 ns. More accurate determination is not possible as $\tau_b$ is at the edge of the experimental accuracy of setting and monitoring the voltage pulses. Importantly, the three other fitting parameters $\alpha_b$, $\alpha_u$, and $\tau_u$ show very little changes with different values of $\tau_b$, as described in Appendix A. In what follows, we set $\tau_b = 1\,\text{ns}$ and fit the experimental data with Eq. (30) using only three fitting parameters: $\tau_u$, $\alpha_u$, and $\alpha_b$.

### B. Uniaxial-Fluctuative geometry fitting

The response of CCN-47 in UF geometry shown in Fig. 6(a) is obtained at the same voltage and temperature as the response in BU geometry, Fig. 5(b). The optic response has two contributions in Eq. (33): the modification of the uniaxial OP and the quenching of director fluctuations. The contribution of the director fluctuations described by Eq. (26) can be simplified for our fitting procedure, because $\tau_d \approx 60\,\text{ms}$ for the cell thickness 4.2 μm, and $\tau_c < 10\,\text{ns}$ for $q_c \approx 1\,\text{nm}^{-1}$. Therefore, the term inside the curly brackets in Eq. (26) is close to unity and

$$\delta\tilde{\varepsilon}_f(t) = A \frac{e^{-S(t)}}{\sqrt{\gamma_{eff}}} \int_0^t \frac{f_E(t')}{\sqrt{(t-t')}} e^{S(t')} dt', \tag{38}$$

where $A$ and $\gamma_{eff}$ are the fitting parameters. Substituting Eq. (37) into Eq. (38), we represent $\delta\tilde{\varepsilon}_f(t)$ by two analytical expressions: switching-on dynamics $\delta\tilde{\varepsilon}_f^{ON}(t_{on} \leq t \leq t_{off})$, and the switching-off dynamics $\delta\tilde{\varepsilon}_f^{OFF}(t > t_{off})$ (see Appendix B for details). The switching-on fluctuations dynamics is

$$\delta\tilde{\varepsilon}_f^{ON}(t_{on} \leq t \leq t_{off}) = A \frac{f_0 e^{-2\nu_1(t-t_{on})}}{\sqrt{\gamma_{eff}}} \left[ a_1^2 \sqrt{\pi \bar{\tau}_f} \; \mathrm{erf}\left(\sqrt{\frac{t-t_{on}}{\bar{\tau}_f}}\right) \right.$$
$$\left. + 2e^{-(t-t_{on})/\bar{\tau}_f} \sum_{i,i'=1}^{3}{}' a_i a_{i'} \frac{\mathrm{D}\left(\sqrt{\lambda_{ii'}(t-t_{on})}\right)}{\sqrt{\lambda_{ii'}}} \right], \tag{39}$$

where $\sum_{i,i'=1}^{3}{}'$ is the sum with the term $i=i'=1$ being excluded; $f_0 = \varepsilon_0 |\Delta\varepsilon| E_0^2$; $\lambda_{ii'} = \nu_i + \nu_{i'} - a_1^2/\tau_f$; $\tau_f = \gamma_{eff}/f_0$; $\bar{\tau}_f = |\lambda_{11}|^{-1} = \tau_f/(a_1^2 - 2\tau_f \nu_1)$ is the characteristic time for the dynamics of fluctuations' quenching; and $\mathrm{D}(z) = e^{-z^2} \int_0^z e^{t^2} dt$ is Dawson's integral; see chapter 7 in [40].

In Eq. (39), the first term, with the error function, provides the main contribution, while the terms with Dawson's integrals describe small corrections caused by the non-square shape of the electric pulse in the NLC. In the case of an ideal square electric pulse, $\tau_a \to \infty$, $\tau_{on} \to 0$, and $\tau_{RC} \to 0$, the terms with Dawson's integrals disappear and $\bar{\tau}_f = \tau_f$.

The switching-off dynamics is

$$\delta\tilde{\varepsilon}_f^{OFF}(t>t_{off}) = A\frac{f_0\sqrt{\pi}}{\sqrt{\gamma_{eff}}}\left\{a_1^2 g e^{-2\nu_1(t_{off}-t_{on})}\sqrt{\overline{\tau}_f}e^{(t-t_{off})/\overline{\tau}_f}\left[\operatorname{erf}\sqrt{\frac{t-t_{on}}{\overline{\tau}_f}}-\operatorname{erf}\sqrt{\frac{t-t_{off}}{\overline{\tau}_f}}\right]+\right.$$

$$+\frac{2}{\sqrt{\pi}}g\sum_{i,i'=1}^{3}\frac{a_i a_{i'}}{\sqrt{\lambda_{ij}}}\left[e^{\frac{-a_1^2}{\tau_f}(t_{off}-t_{on})}D\left(\sqrt{\lambda_{ij}(t-t_{on})}\right)-e^{-(\nu_i+\nu_{i'})(t_{off}-t_{on})}D\left(\sqrt{\lambda_{ij}(t-t_{off})}\right)\right]+ \quad (40)$$

$$\left.+\frac{f_E^{ON}(t_{off})}{f_0}\frac{2}{\sqrt{\pi}}\sum_{j,j'=1}^{2}\frac{b_j b_{j'}}{\sqrt{\mu_j+\mu_{j'}}}D\left(\sqrt{(\mu_j+\mu_{j'})(t-t_{off})}\right)\right\},$$

where $g = \exp\left[-\frac{f_E^{ON}(t_{off})}{\gamma_{eff}}\sum_{j,j'=1}^{2}\frac{b_j b_{j'}}{\mu_j+\mu_{j'}}\right]$.

Fitting the experimental data with the corresponding Eqs. (12), (39) and (40) reveals that the characteristic time of the fastest process is about 30 ns, and there is no process with the characteristic time on the order of 1 ns, which we observe in BU geometry, Fig. 5(b). Therefore, the UF experiment proves our earlier assignment that the relatively slow (30 ns) process in BU geometry is related to the modification of the uniaxial OP, and the fast nanosecond process is caused by the induced biaxial OP.

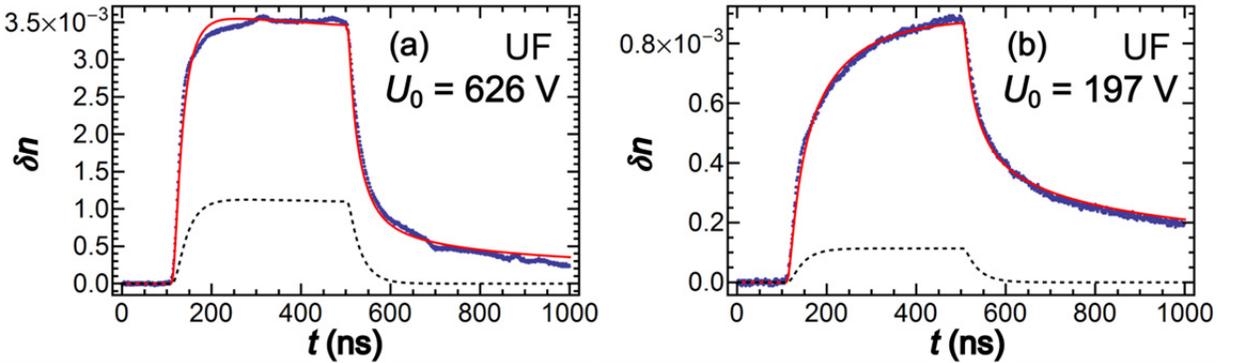

**FIG. 6.** (Color online) Optic response measured in UF geometry at 46°C. Uniaxial component $\delta\tilde{\varepsilon}_u(t)$ parameters $\alpha_u$ and $\tau_u$ obtained from BU geometry at voltage $U_0$ were used to fit UF geometry data and to obtain $A$ and $\gamma_{eff}$. (a) $\alpha_u = 9.5\times 10^{-20}$ m$^2$/V$^2$ and $\tau_u = 28$ ns for the applied voltage pulse $U_0 = 626$ V yield parameters $A = 1.7\,\mu\text{s}(\text{m/kg})^{1/2}$ and $\gamma_{eff} = 25$ mPa s. (b) $\alpha_u = 9.6\times 10^{-20}$ m$^2$/V$^2$ and $\tau_u = 30$ ns for $U_0 = 197$ V pulse yield $A = 1.7\,\mu\text{s}(\text{m/kg})^{1/2}$ and $\gamma_{eff} = 15$ mPa s. The experimental points are fitted with our model (solid red line), and the dashed line is the uniaxial contribution, $\delta\tilde{\varepsilon}_u(t)$, obtained from BU geometry.

The reliable fitting of the uniaxial and fluctuations contributions with Eqs. (12), (39), and (40) might be challenging, especially for higher electric fields, when the characteristic times $\bar{\tau}_f$ and $\tau_u$ are of the same order. On the other hand, $\tau_b$ and $\tau_u$ are more than one order of magnitude different, and fitting BU geometry allows us to obtain the biaxial and uniaxial contributions with high accuracy. Therefore, we separate the uniaxial contribution from the experimental data in UF geometry using the corresponding fitting parameters $\alpha_u$ and $\tau_u$ obtained from BU geometry for the same temperature and voltage pulse. Then we fit the remaining part corresponding to the director fluctuations with Eqs. (39) and (40). Although we use only two fitting parameters $A$ and $\gamma_{eff}$, the experimental data fit for UF geometry is encouraging, both for higher electric fields when the optic response is faster, Fig. 6(a), and for lower fields when the response is slower, Fig. 6(b).

### C. Normal geometry

Using an arbitrary direction of the probing beam propagation in our experimental system, one can obtain a linear combination of two independent experimental sets of data, Eq. (29). More specifically, the optic response in N geometry can be presented as the linear combination of respective responses in BU and UF geometries. In order to validate the two experimental sets of data taken in BU and UF geometries, we perform an experiment in N geometry.

With a probing beam impinging normal on the substrates, N geometry contains the contributions of all three processes, Eq. (34): the field-enhanced uniaxial OP, field-induced biaxial OP, and the quenching of director fluctuations. Equations (30), (33), and (34) show that the linear combination of the optic responses in BU, UF, and N geometries, expressed as

$$\delta n_0(t) = \delta n_N(t) - \frac{\sqrt{n_e^2 + n_o^2}}{n_o^2/n_e + n_o + n_e}\left[\delta n_{BU}(t) - \frac{n_e^2 - n_o^2}{\sqrt{n_e^2 + n_o^2}}(\bar{\beta} - \bar{\beta}_0)\right]$$
$$- \frac{\sqrt{2}n_o\sqrt{2n_e^2 - n_o^2}}{n_e\left(\sqrt{2n_e^2 - n_o^2} + 2n_o\right)}\delta n_{UF}(t)$$
(41)

should be zero. This quantity can be used as an estimate of the experimental error. In all our experiments, the field-induced phase difference, $\delta n_0(t)$, described in Eq. (41), deviates from zero by no more than $1.4\times 10^{-4}$ (except at the moments of time corresponding to the front and rear edges of the voltage pulse), Fig.7.

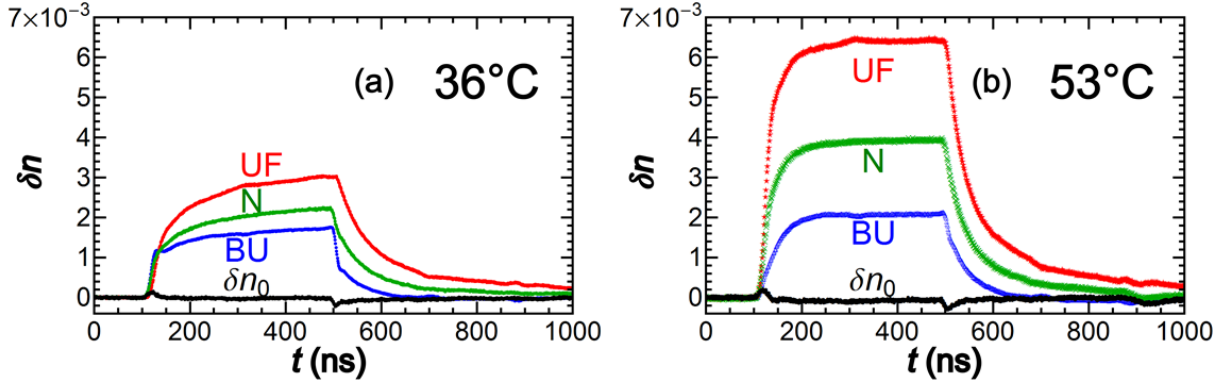

FIG. 7. (Color online) Optic responses measured in geometries BU, UF, and N at (a) 36°C and (b) 53°C. The lowest black curve corresponds to $\delta n_0(t)$ defined in Eq. (41). Applied voltage pulse $U_0 = 626$ V.

## V.  DISCUSSION
### A. Biaxial-Uniaxial geometry

The experimental data follow our model fairly well, Figs. 5(b) and 13(a). In particular, at the temperatures $T$ = 31°C, 46°C, and 49°C, Fig. 8, that are far from the nematic-to-isotropic ($T_{NI}$ = 56.5°C) phase transition, the fitting parameters, namely, the biaxial $\alpha_b$ and uniaxial $\alpha_u$ susceptibilities and the characteristic uniaxial time $\tau_u$, do not depend on the electric field, as expected, see Eq. (12). Close to $T_{NI}$, at $T$ = 54°C, $\alpha_u$ and $\tau_u$ decrease, while $\alpha_b$ increases with the electric field. Such a behavior in the pretransitional region might be attributed to the following factors. First, we restrict our model by the second-order term of the free energy density expansion, Eq. (4). One can expect that near the $T_{NI}$, the higher-order terms should be taken into account. Second, while our model describes the NEMOP effect through four OPs, Eq. (11), we fit experimental data with the assumption of only two OPs being significant ($R_{00}$ and $R_{22}$), Eq. (12).

The temperature dependences of $\alpha_u$ and $\tau_u$, shown on Fig. 9, are obtained for $E_0 = 74$ V/μm. Such a field is not very strong, yet the induced optic response is sufficiently large to provide reasonable accuracy.

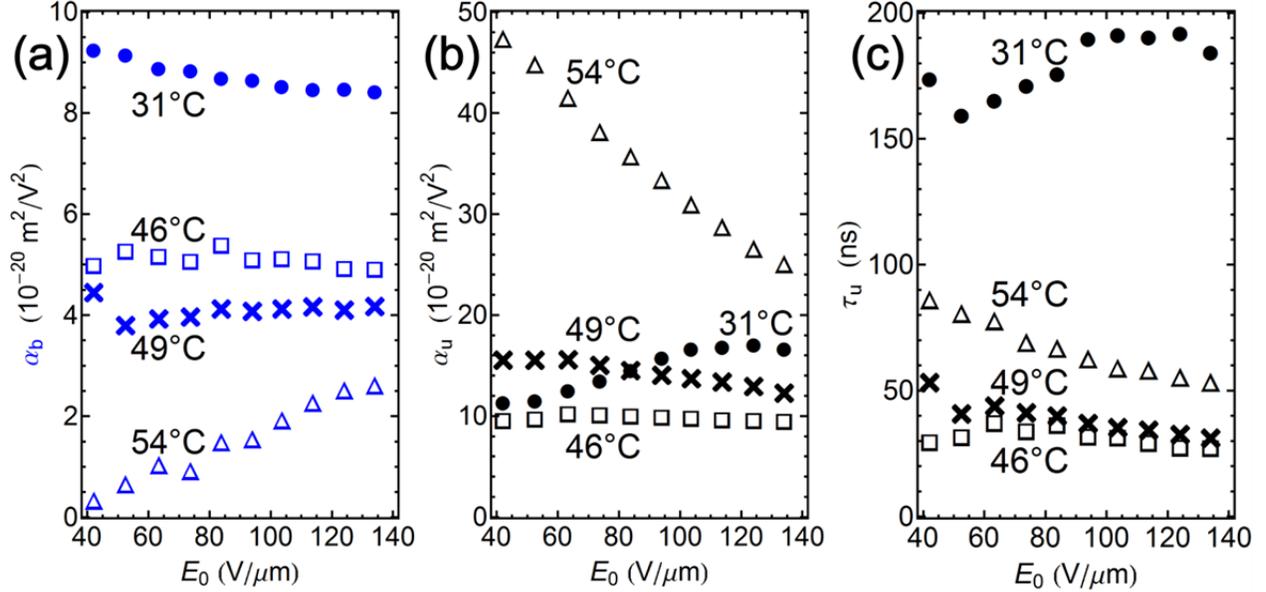

**FIG. 8.** (Color online) Electric field dependence of (a) biaxial $\alpha_b$, (b) uniaxial $\alpha_u$ susceptibilities, and (c) uniaxial time $\tau_u$ at different temperatures: 31°C (●), 46°C (□), 49°C (✗), and 54°C (Δ).

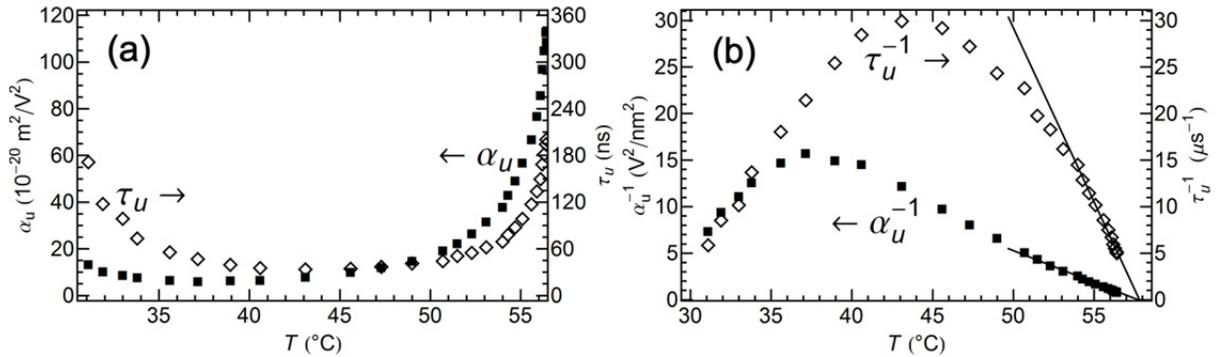

**FIG. 9.** Temperature dependences of (a) uniaxial susceptibility $\alpha_u$ (■), and uniaxial characteristic time $\tau_u$ (◇) measured at $E_0 = 74$ V/μm; and (b) their reciprocal values $\alpha_u^{-1}$ and $\tau_u^{-1}$ fitted with straight lines.

When the temperature approaches $T_{NI}$, both the uniaxial susceptibility $\alpha_u$ and relaxation time $\tau_u$ increase, Fig. 9(a). In the theory, both quantities are inversely proportional to $\partial^2 f_m / \partial R_{jk} \partial R_{jk'}$, see Eq. (4), i.e., $\alpha_u \propto 1/M_{00,00}$ and $\tau_u \propto 1/M_{00,00}$. The experimentally observed increase of $\alpha_u$ and $\tau_u$ is, thus, explained by the flattening of the free energy density profile as a function of $R_{00}$ near the phase transition temperature. Therefore, the experimental behavior of $\alpha_u$ and $\tau_u$ is consistent with the Landau-Khalatnikov description close to the phase transition [34].

The reciprocal quantities $1/\alpha_u$ and $1/\tau_u$ demonstrate a quasi-linear behavior at both low and high temperatures of the nematic range, Fig 9(b). Close to $T_{NI}$, this behavior could be explained by the Landau-de Gennes theory for the nematic phase, where $M_{00,00}$ has a quasi-linear temperature dependence, and adopts a zero value at the absolute temperature limit $T^{**}$ of overheating of the nematic phase, Fig. 9(b).

At the lower temperature limit of the nematic phase, the value of $\alpha_u$ slightly increases, Fig. 9(b), which could be attributed to the formation of fluctuative smectic clusters near the nematic-to-smectic phase transition, which is enhanced by the electric field. Clusters might also explain the increase of the response time $\tau_u$ at the low temperatures.

The biaxial susceptibility $\alpha_b$ shows a well-pronounced increase as the temperature is lowered, Fig. 10(a), which can be explained in the following way. In our model, $\alpha_b$ is proportional to $M_{22,22}^{-1}$, Eq. (13). According to the Landau theory, the biaxial second-order coefficient $M_{22,22}$ in the uniaxial phase, Eq. (4), has to go to zero at the temperature $T_{ub}$ of the uniaxial-biaxial nematic phase transition, and this dependence is linear $M_{22,22} \propto (T - T_{ub})$. Therefore, one can expect that $\alpha_b^{-1} \propto (T - T_{ub})$ and the experimental data show such a linear dependence for temperatures far below $T_{NI}$, Fig. 10(b). The slope of the linear temperature dependence of $\alpha_b^{-1}$ shows that the hypothetical uniaxial-to-biaxial nematic phase transition temperature is $T_{ub} = 5°C$, Fig. 10(b). This temperature is well below the uniaxial-to-smectic A transition temperature $T_{NA} = 30°C$ observed for CCN-47. Thus, the molecular structure of CCN-47 is not conducive for the search of a biaxial nematic phase. On a general note, the temperature dependence of $\alpha_b$ can serve as an indicator of how close a uniaxial nematic material might be to forming a biaxial nematic phase in absence of the external electric field.

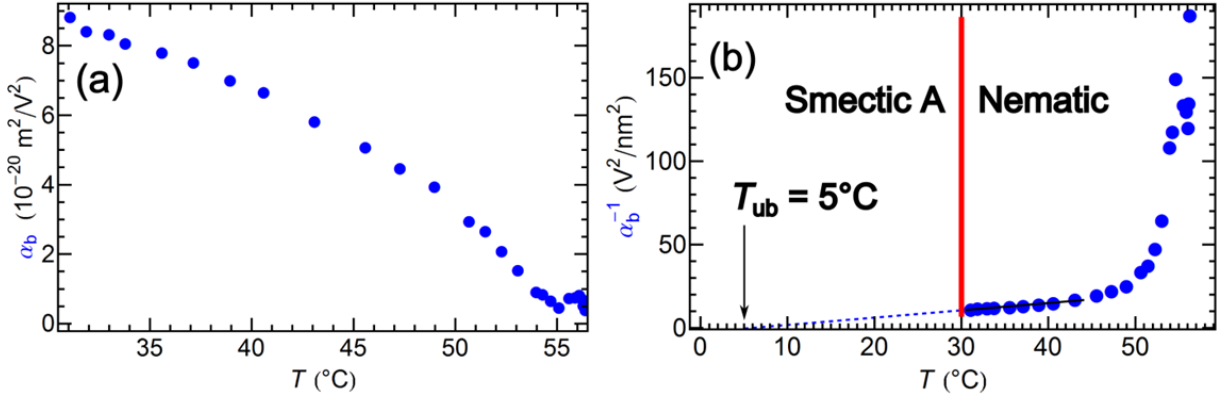

**FIG. 10.** (Color online) Temperature dependences of (a) biaxial susceptibility $\alpha_b$ (●) and (b) its reciprocal $\alpha_b^{-1}$ fitted with a straight line.

### B. Uniaxial-Fluctuative geometry

This geometry offers a convenient way for analyzing the nanosecond dynamics of the quenching of director fluctuations, because the biaxial contribution is absent and the uniaxial contribution in Eq. (33) can be separated from the fluctuative contribution since the vaues of $\alpha_u$ and $\tau_u$ are already known from the fit of the experimental data in BU geometry. The electric-field dependences of the fitting parameters $A$ and $\gamma_{eff}$ for several temperatures are shown in Fig. 11. As expected, the amplitude coefficient $A$, describing the changes in the optic tensor caused by the quenching of director fluctuations, Eq. (26), remains almost field-independent and increases with temperature, Fig. 11(a) and Fig. 12. However, the value of $A$ is about two times bigger than the value expected from its definition in Eq. (26), calculated with the known elastic constants [42] and the measured $n_e = 1.50$ and $n_o = 1.47$.

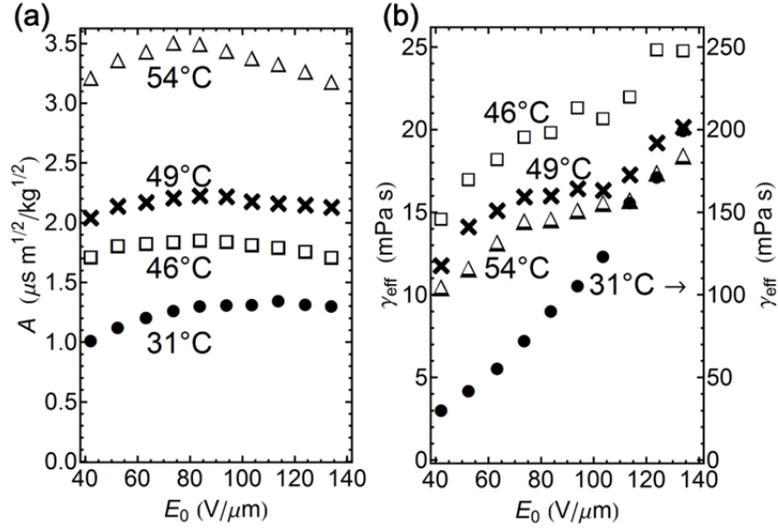

**FIG. 11.** Fitting parameters (a) $A$ and (b) $\gamma_{eff}$ obtained from experimental data at 31°C (●), 46°C (□), 49°C (✕), and 54°C (▲).

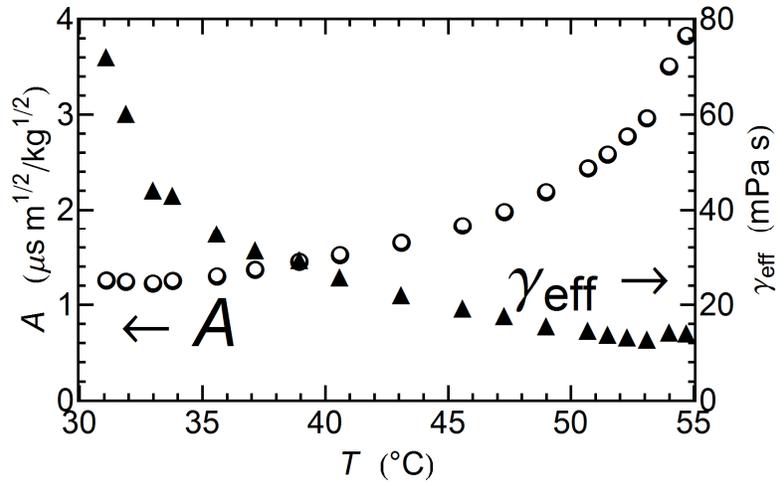

**FIG. 12.** Temperature dependence of $A$ (○) and $\gamma_{eff}$ (▲) at $E_0 = 74$ V/μm.

The obtained effective viscosity $\gamma_{eff}$ demonstrates a weak monotonous increase with electric field, Fig. 11(b) and is slightly smaller than the macroscopic viscosities of CCN-47 homologue compounds and their mixtures [43]. As expected, in the nematic phase, $\gamma_{eff}$ increases with a decrease in temperature, Fig. 12. The increase is especially pronounced near the transition to the smectic A phase. The latter can be attributed to the pre-transitional phenomena such as fluctuative cybotactic smectic clusters.

## VI. CONCLUSION

In this work, we explored both theoretically and experimentally the electro-optic response of an NLC cell in which the electric field does not cause director reorientation. We demonstrated three mechanisms contributing to the field-induced change of optical birefringence: nanosecond electric modification of (a) biaxial and (b) uniaxial OPs and (c) quenching of the director fluctuations. Our observations reveal that these mechanisms have different characteristic times. For CCN-47, these times are (a) less than 2 nanoseconds for the biaxial NEMOP, (b) tens of nanoseconds for the uniaxial NEMOP, and (c) a wide range of characteristic times from tens of nanoseconds to milliseconds for the quenching of director fluctuations.

We developed a model of the NEMOP effect using two uniaxial and two biaxial nematic OPs. Their dynamics are described by two uniaxial and two biaxial modes, Eq. (5). We used a simplified two-mode version of the model to fit our experimental data for CCN-47; the uniaxial OP of the long molecular axes and the biaxial OP of the short molecular axes appear to be the dominant OPs for this material.

We describe the dynamics of director fluctuations using the macroscopic viscoelastic approach, Eq. (20), with Frank-Oseen elastic energy in splay-twist one-constant approximation, $K_1 = K_2$, and with a constant effective viscosity. Within these approximations, we derived the contribution for the quenching of director fluctuations to the field-induced modifications of the optic tensor, Eq. (26).

Experimentally, we determine the field-induced changes of the effective birefringence $\delta n_{eff}$, which contains the uniaxial $\delta\tilde{\varepsilon}_u$, biaxial $\delta\tilde{\varepsilon}_b$, and fluctuational $\delta\tilde{\varepsilon}_f$ contributions, Eq. (29). In order to separate these contributions, we used the so-called biaxial-uniaxial (BU) and uniaxial-fluctuative (UF) geometries, in which one of the three contributions is nullified. We also independently validated the separation of different mechanisms by measuring the optic response in normal incidence (N) geometry, Fig. 7.

In BU geometry, with no contribution from the fluctuations quenching, the dynamics of electro-optic response develops over timescales of nanoseconds and is well described by two different characteristic times $\tau_u$ (tens of nanoseconds), and $\tau_b$ (about two nanoseconds or less). We associate these characteristic times with the uniaxial and biaxial modifications of the optic tensor, respectively, see Eqs. (30) and (12). The assignment of the fastest relaxation time $\tau_b$ to the biaxial modification is justified by the measurements in UF geometry, in which the nanosecond relaxation is absent. The biaxial susceptibility shows a strong temperature dependence at low temperatures, $\alpha_b \propto (T-T_{ub})^{-1}$, which indicates a possible phase transition from the uniaxial to the biaxial nematic phase in a field-free state, at some temperature $T_{ub}$. The extrapolated value is $T_{ub} = 5°C$, much lower than the temperature 30°C of the actual phase transition from the uniaxial nematic to the smectic A phase. Therefore, in the explored material CCN-47, the hypothetical biaxial nematic state is suppressed by the occurrence of the smectic A phase. A similar test can be used to find $T_{ub}$ in other materials, in order to facilitate the search for potential biaxial nematics.

UF geometry provides interesting information about the behavior of director fluctuations on nanoseconds' timescales. In this geometry, the biaxial modifications in the optic tensor $\delta\tilde{\varepsilon}_b$ are eliminated and the uniaxial changes can be evaluated by employing the values of parameters $\alpha_u$ and $\tau_u$ obtained from the 'slow' component of the BU response. The remaining changes $\delta\tilde{\varepsilon}_f$ in the optic tensor can be attributed to the quenching of director fluctuations. The director fluctuations model provides a good fit to the experimental optic response, Fig. 6. As expected, the amplitude of director fluctuations grows with temperature, while the effective viscosity decreases with temperature, Fig. 12. The amplitude coefficient $A$ does not depend on the electric field but is bigger than theoretically expected, Fig. 11(a), what can be attributed to the simplifying assumptions of the theory. The most intriguing feature is that the effective viscosity increases with the field, Fig. 11(b), thus, possibly indicating that the classic viscoelastic theory with constant material parameters might approach its limit of validity when applied to the nanoseconds dynamics in strong electric fields.

The presented NEMOP effect should be distinguished from the classic Kerr effect. The Kerr effect consists in field-induced birefringence emerging in the otherwise isotropic fluid. It is an essentially uniaxial effect, with the induced optic axis being always parallel to the applied field. The Kerr effect can be observed in non-mesogenic fluids [44-46] and in the isotropic phase of mesogenic compounds [47-52]. In the first case, the effect is practically temperature independent, while in the second case, it shows a strong enhancement near the isotropic-to-nematic phase transition [50, 52]. In comparison, the NEMOP response of CCN-47 with a negative dielectric anisotropy features both uniaxial and biaxial optical changes. The biaxial changes are faster than the uniaxial changes at the same temperature and in the same electric field, as discussed above. Similarly to the case of electro-optic effects in uniaxial and biaxial nematics [53], one could expect that the biaxial part of NEMOP would be generally faster than the uniaxial part. It is also expected that the relative contributions of the biaxial and uniaxial changes, the amplitude and relaxation times of these changes would be strongly dependent on the molecular structure, as the NEMOP effect is essentially a molecular-scale phenomenon. Indeed, our recent results [13] demonstrate that different mesogenic materials show very different amplitudes of the field-induced NEMOP birefringence that exceed the data presented for CCN-47 by at least one order of magnitude.

From the fundamental point of view, NEMOP represents an opportunity to analyze the complex uniaxial-biaxial response of the orientationally ordered medium to the applied electric field at the scale of nanoseconds. In this work, we explored only one material. Further studies should expand to materials with different molecular structures and material parameters. For instance, the NEMOP effect can be observed not only in materials with a negative dielectric anisotropy, as is the case of CCN-47, but also in materials with positive dielectric anisotropy. It would be of interest to compare the parameters of NEMOP effect to the parameters of the Kerr effect in the isotropic phase of the same compound. These studies would shed some light on which mode of optic response would be the most beneficial for the nanosecond electro-optic applications.

**ACKNOWLEDGEMENTS**

We thank M. Groom and C. Culbreath for their help with the experimental setup. The work was supported by DOE grant DE-FG02-06ER 46331 (measurements of the optic response and theory), NSF DMR-1410378, State of Ohio through Ohio Development Services Agency and Ohio Third Frontier grant TECG20140133 (analysis of the field and temperature dependences of the optic response). B.X.L. acknowledges China Scholarship Council and Jiangsu Innovation Program for Graduate Education grant CXLX13_155 support. The content of this publication reflects the views of the authors and does not purport to reflect the views of the Ohio Development Services Agency and/or that of the State of Ohio.

**APPENDIX A: FITTING PROCEDURE FOR BIAXIAL-UNIAXIAL GEOMETRY**

In this Appendix, we explain the procedure to fit the experimental data obtained in BU geometry. There are three processes that are relevant in the dynamics of optic response in this geometry, namely, director reorientation associated with the finite pretilt, biaxial and uniaxial changes of the OPs.

The slowest one is the dynamics of the pretilt angle $\bar{\beta}(t)$, described by Eqs. (31) and (32). When the field is switched on, the characteristic time $\tau_{on}^{F} \approx \gamma_{1}/(\varepsilon_{0}|\Delta\varepsilon|E^{2})$ of the pretilt dynamics with $\gamma_{1} \approx 0.1$ Pa s being the rotational viscosity and $E \approx 2 \times 10^{8}$ V/m being the typical electric field, is about 100 ns, which is longer than the rate of uniaxial and biaxial changes, $\tau_{u} \sim 30$ ns and $\tau_{b} < 2$ ns. When the electric field is switched off at $t = t_{off}$, the relaxation time of the pretilt angle becomes even longer, $\tau_{off}^{F} \approx \gamma_{1}d^{2}/(\pi^{2}K_{1}) \sim 10$ ms. At the scale of nanoseconds relevant to our experiments, this extremely slow relaxation yields a practically time-independent contribution to the overall optical signal that reveals itself in Fig. 13(a) as a negative-valued 'tail' in the time dependence of $\delta n$ (see also Figs. 3c, 5b). Since the uniaxial and biaxial modifications relax much faster than the pretilt angle, we use the optic signal measured at $t > t_{off} + 500$ ns to determine the value of $\bar{\beta}(t > t_{off})$; the value of $\bar{\beta}_{0}$ follows from Eq. (31). Note that the overall effect of $\bar{\beta}(t)$ is small, contributing less than 5% to the optic response.

After the exclusion of the pretilt angle contribution, the remaining dynamics is associated with the uniaxial and biaxial changes of the OPs that occur on short timescales, (1-100) ns. We fit the experimental data with Eq. (30) in which $\bar{\beta}(t)$ is defined as explained above. The fitting is performed through minimization of the residuals function

$$\text{var} = \frac{1}{N-4}\sum_{i=1}^{N}\left[\delta n(t_{i}) - \delta n_{BU}(t_{i}, \alpha_{u}, \alpha_{b}, \tau_{u}, \tau_{b})\right]^{2}, \tag{A1}$$

where $N$ is the number of experimental data points $\{t_{i}, \delta n(t_{i})\}$ and $\delta n_{BU}$ is the fitting function as defined in Eq. (30).

The fitting clearly reveals two different relaxation processes with substantially different relaxation times: $\tau_{u}$ in the range of tens of nanoseconds and $\tau_{b}$ in the range of nanoseconds. For example, the optic response to the voltage $U_0 = 626$ V, yields $\tau_b = 1.8$ ns and $\tau_u = 31$ ns, Fig. 13(a). As long as $\tau_b$ is less than 2ns, the fitting produces practically the same values of the three other parameters, $\alpha_u$, $\alpha_b$, and $\tau_u$, Fig. 13(b),(c),(d).

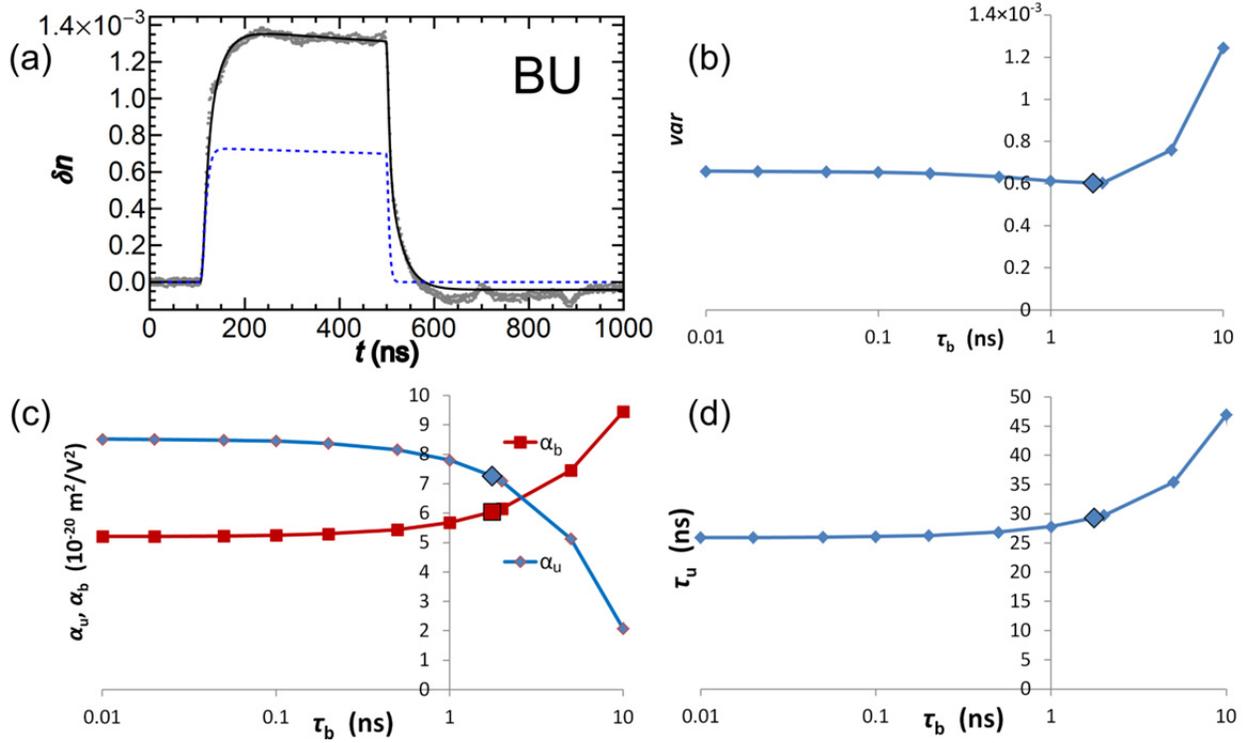

**FIG. 13.** (Color online) (a) Optic response at $T = 43°C$ (gray dots) fitted with Eq. (30) for one uniaxial and one biaxial mode, $\tau_b = 1.76$ ns, $\tau_u = 31$ ns, $\alpha_b = 5.8 \times 10^{-20}$ m$^2$/V$^2$, and $\alpha_u = 8.0 \times 10^{-20}$ m$^2$/V$^2$ (solid black line). The blue dashed line is the biaxial contribution. (b) Dependence of the residuals function on the preselected value of $\tau_b$, obtained from the fitting of the optic response at $T = 43°C$, $U_0 = 626$ V with Eq. (30). Dependence of the fitted values of $\alpha_u$, $\alpha_b$ (c) and $\tau_u$ (d) on the preselected value of $\tau_b$. The big marker on the plots corresponds to $\tau_b = 1.76$ ns, obtained as a free fitting parameter.

# APPENDIX B: ANALYTIC DESCRIPTION OF THE DYNAMICS OF DIRECTOR FLUCTUATIONS QUENCHING

In this Appendix, we derive an expression for the dynamics of the fluctuative contribution described by Eq. (38). To simplify derivation we set $t_{on} = 0$. The function $f_E(t) = \varepsilon_0 |\Delta\varepsilon| E^2(t)$ reads from Eq. (37) as

$$f_E^{ON}(t \leq t_{off}) = f_0 \sum_{i,i'=1}^{3} a_i a_{i'} e^{-(\nu_i + \nu_{i'})t},$$

$$f_E^{OFF}(t > t_{off}) = f_E^{ON}(t_{off}) \sum_{j,j'=1}^{2} b_j b_{j'} e^{-(\mu_j + \mu_{j'})(t-t_{off})},$$
(B1)

where $f_0 = \varepsilon_0 |\Delta\varepsilon| E_0^2$, $E_0 = U_0 \varepsilon_P / (\varepsilon_\perp d_P + \varepsilon_P d)$ is the characteristic amplitude of the electric field inside the NLC, Eq. (37), $i = 1, 2, 3$, see Table I of the main text, and $j = 1, 2$, see Table II.

For the switching-on process, $t \leq t_{off}$, $S^{ON}(t \leq t_{off}) = \frac{1}{\gamma_{eff}} \int_0^t f_E^{ON}(t')dt'$, therefore, $\exp[S^{ON}(t)]$ in Eq. (38) can be presented in a form $\exp[S^{ON}(t)] = \prod_{i,i'=1}^{3} P_{ii'}(t)$, where $P_{ii'}(t) = \exp\left[\frac{a_i a_{i'}}{\tau_f (\nu_i + \nu_{i'})} \left(1 - e^{-(\nu_i + \nu_{i'})t}\right)\right]$ and $\tau_f = \gamma_{eff}/f_0$. One can see from Table I, that $|a_i| \sim 1$ and $\nu_i$ satisfy the conditions $\nu_1 \tau_f \ll 1$ and $\nu_i \tau_f \gg 1$ for $i = 2, 3$. Thus, the exponential term in parentheses can be expanded for $P_{11}(t) = \exp(a_1^2 t / \tau_f)$ and neglected for all other terms $P_{ii'} = \exp\left[\frac{a_i a_{i'}}{\tau_f (\nu_i + \nu_{i'})}\right]$. Therefore, Eq. (38) can be presented as

$$\delta\tilde{\varepsilon}_f^{ON}(t \leq t_{off}) = A\sqrt{\pi} \frac{f_0}{\sqrt{\gamma_{eff}}} e^{-\frac{a_1^2}{\tau_f}t} \sum_{i,i'=1}^{3} a_i a_{i'} I(\lambda_{ii'}, 0, t),$$
(B2)

where $\lambda_{ii'} = \nu_i + \nu_{i'} - \frac{a_1^2}{\tau_f}$ and $I(\lambda, t_0, t) = \int_{t_0}^{t} \frac{e^{-\lambda t'}}{\sqrt{t - t'}} dt'$. The integral $I(\lambda, t_0, t)$ yields either the error function, or Dawson's integral function, see, e.g., chapters 5 and 7 of Ref. [40]:

$$I(\lambda,t_0,t) = \begin{cases} \sqrt{\pi}\,\dfrac{e^{|\lambda|t}}{\sqrt{|\lambda|}}\,\text{erf}\sqrt{|\lambda|(t-t_0)} & \text{if } \lambda < 0, \\ 2\dfrac{e^{-\lambda t_0}}{\sqrt{\lambda}}\,\text{D}\left(\sqrt{\lambda(t-t_0)}\right) & \text{if } \lambda > 0. \end{cases} \quad (B3)$$

One can see from Table I, that $\lambda_{11} < 0$ and $\lambda_{ii'} > 0$ for all other cases; thus, Eq. (38) for the switch-on dynamics, $t \leq t_{off}$, becomes

$$\delta\tilde{\varepsilon}_f^{ON}(t \leq t_{off}) = A\frac{f_0}{\sqrt{\gamma_{eff}}}e^{-\frac{a_1^2}{\tau_f}t}\left[\sqrt{\pi}\frac{a_1^2 e^{|\lambda_{11}|t}}{\sqrt{|\lambda_{11}|}}\text{erf}\sqrt{|\lambda_{11}|t} + 2\sum_{i,i'=1}^{3}{}'\frac{a_i a_{i'}}{\sqrt{\lambda_{ii'}}}\text{D}\left(\sqrt{\lambda_{ii'}t}\right)\right], \quad (B4)$$

where $\sum_{i,i'=1}^{3}{}'$ is the sum with the term $i = i' = 1$ being excluded. Equation (B4) is presented as Eq. (39) in the main text.

For the switching-off process, $t > t_{off}$, we can split $S(t > t_{off})$ into two parts $S(t > t_{off}) = S^{ON}(t_{off}) + S^{OFF}(t > t_{off})$, where

$$S^{OFF}(t > t_{off}) = \frac{1}{\gamma_{eff}}\int_{t_{off}}^{t} f_E^{OFF}(t')dt' = \frac{f_E^{ON}(t_{off})}{\gamma_{eff}}\sum_{j,j'=1}^{2}\frac{b_j b_{j'}}{(\mu_j + \mu_{j'})}\left(1 - e^{-(\mu_j + \mu_{j'})(t - t_{off})}\right). \quad (B5)$$

Thus, Eq. (38) is also divided into two parts:

$$\delta\tilde{\varepsilon}_f^{OFF}(t > t_{off}) = \frac{A}{\sqrt{\gamma_{eff}}}e^{-S^{OFF}(t)}\left[e^{-S^{ON}(t_{off})}\int_0^{t_{off}}\frac{f_E^{ON}(t')e^{S^{ON}(t')}}{\sqrt{t-t'}}dt' + \int_{t_{off}}^{t}\frac{f_E^{OFF}(t')e^{S^{OFF}(t')}}{\sqrt{t-t'}}dt'\right]. \quad (B6)$$

During the switching-off process, $\mu_j \tau_f \gg 1$, thus we can neglect the exponential term in Eq. (B5), so that $\exp[-S^{OFF}(t)] \approx \exp\left[-\dfrac{f_E^{ON}(t_{off})}{\gamma_{eff}}\sum_{j,j'=1}^{2}\dfrac{b_j b_{j'}}{\mu_j + \mu_{j'}}\right] = g \approx 1$. Therefore,

$$\delta\tilde{\varepsilon}_f^{OFF}(t > t_{off}) = \frac{A}{\sqrt{\gamma_{eff}}}\left[gf_0 e^{-\frac{a_1^2}{\tau_f}t_{off}}\sum_{i,i'=1}^{3}a_i a_{i'}\int_0^{t_{off}}\frac{e^{-\lambda_{ii'}t'}}{\sqrt{t-t'}}dt' + f_E^{ON}(t_{off})\sum_{j,j'=1}^{2}b_j b_{j'}\int_{t_{off}}^{t}\frac{e^{-(\mu_j + \mu_{j'})(t' - t_{off})}}{\sqrt{t-t'}}dt'\right]. \quad (B7)$$

Representing the integrals in the first sum of Eq. (B7) as $\int_0^{t_{off}}dt' = \int_0^{t}dt' - \int_{t_{off}}^{t}dt'$ and using Eq. (B3), we obtain Eq. (40).

*References*

[1]     P. G. de Gennes and J. Prost, *The Physics of Liquid Crystals* (Clarendon Press, Oxford, 1993).

[2]     C. P. Fan and M. J. Stephen, *Phys. Rev. Lett.* **25**, 500 (1970).

[3]     P. Palffy-Muhoray and D. A. Dunmur, *Mol. Cryst. Liq. Cryst.* **97**, 337 (1983).

[4]     A. J. Nicastro and P. H. Keyes, *Phys. Rev. A* **30**, 3156 (1984).

[5]     E. F. Gramsbergen, L. Longa, and W. H. de Jeu, *Phys. Rep.* **135**, 195 (1986).

[6]     I. Lelidis, M. Nobili, and G. Durand, *Phys. Rev. E* **48**, 3818 (1993).

[7]     I. Lelidis and G. Durand, *Phys. Rev. E* **48**, 3822 (1993).

[8]     J. A. Olivares, S. Stojadinovic, T. Dingemans, S. Sprunt, and A. Jákli, *Phys. Rev. E* **68**, 041704 (2003).

[9]     R. Stannarius, A. Eremin, M. G. Tamba, G. Pelzl, and W. Weissflog, *Phys. Rev. E* **76**, 061704 (2007).

[10]    M. Nagaraj, Y. P. Panarin, U. Manna, J. K. Vij, C. Keith, and C. Tschierske, *Appl. Phys. Lett.* **96**, 011106 (2010).

[11]    V. Borshch, S. V. Shiyanovskii, and O. D. Lavrentovich, *Mol. Cryst. Liq. Cryst.* **559**, 97 (2012).

[12]    V. Borshch, S. V. Shiyanovskii, and O. D. Lavrentovich, *Phys. Rev. Lett.* **111**, 107802 (2013).

[13]    B.-X. Li, V. Borshch, S. V. Shiyanovskii, S.-B. Liu, and O. D. Lavrentovich, *Appl. Phys. Lett.* **104**, 201105 (2014).

[14]    M. Kleman and O. D. Lavrentovich, *Soft Matter Physics: An Introduction* (Springer, New York, 2003).

[15]    P. G. de Gennes, *C.R. Acad. Sci. Paris* **266B**, 15 (1968).

[16]    Groupe d'Etudes des Cristaux Liquides (Orsay), *J. Chem. Phys.* **51**, 816 (1969).

[17]    J. L. Martin and G. Durand, *Solid State Commun.* **10**, 815 (1972).

[18]    Y. Poggi and J. C. Filippini, *Phys. Rev. Lett.* **39**, 150 (1977).

[19]    T. E. Faber, *Proc. R. Soc. London. Ser. A.* **353**, 247 (1977).

[20]    B. Malraison, Y. Poggi, and E. Guyon, *Phys. Rev. A* **21**, 1012 (1980).

[21]    M. Warner, *Mol. Phys.* **52**, 677 (1984).

[22]   A. Seppen, G. Maret, A. Jansen, P. Wyder, J. Janssen, and W. de Jeu, in *Biophysical effects of steady magnetic fields: proceedings of the workshop, Les Houches*, (Springer, 1986), p. 18.

[23]   D. A. Dunmur and P. Palffy-Muhoray, *J. Phys. Chem.* **92**, 1406 (1988).

[24]   T. E. Faber, *Liq. Cryst.* **9**, 95 (1991).

[25]   B. J. Gertner and K. Lindenberg, *J. Chem. Phys.* **94**, 5143 (1991).

[26]   M. J. Freiser, *Phys. Rev. Lett.* **24**, 1041 (1970).

[27]   J. P. Straley, *Phys. Rev. A* **10**, 1881 (1974).

[28]   C. Zannoni, in *The Molecular Physics of Liquid Crystals*, edited by G. R. Luckhurst and G. W. Gray (Academic Press, 1979).

[29]   G. R. Luckhurst, *Liq. Cryst.* **36**, 1295 (2009).

[30]   D. A. Varshalovich, A. N. Moskalev, and V. K. Khersonskiĭ, *Quantum Theory of Angular Momentum: Irreducible Tensors, Spherical Harmonics, Vector Coupling Coefficients, 3 Nj Symbols* (World Scientific Publishing Company, Incorporated, 1988).

[31]   A. M. Sonnet, E. G. Virga, and G. E. Durand, *Phys. Rev. E* **67**, 061701 (2003).

[32]   S. V. Shiyanovskii, *Phys. Rev. E* **87** (2013).

[33]   Y. Yin, S. V. Shiyanovskii, and O. D. Lavrentovich, *J. Appl. Phys.* **100**, 024906 (2006).

[34]   L. D. Landau and I. M. Khalatnikov, *Dokl. Akad. Nauk SSSR* **96**, 469 (1954).

[35]   S. Hess, *Z. Naturforsch.* **31a**, 1507 (1976).

[36]   P. D. Olmsted and P. M. Goldbart, *Phys. Rev. A* **46**, 4966 (1992).

[37]   R. Berardi, L. Muccioli, S. Orlandi, M. Ricci, and C. Zannoni, *J. Phys. Condens. Matter* **20**, 463101 (2008).

[38]   S. V. Shiyanovskii, *Ukr. J. Phys.* **26**, 137 (1981).

[39]   B. Y. Zel'dovich and N. V. Tabiryan, *Sov. Phys. JETP* **54**, 922 (1981).

[40]   M. Abramowitz and I. A. Stegun, *Handbook of mathematical functions: with formulas, graphs, and mathematical tables* (Dover Publications, 1964).

[41]   R. N. Thurston, G. D. Boyd, and D. C. Senft, *J. Appl. Phys.* **55**, 3846 (1984).

[42]   H. G. Walton, *Mol. Cryst. Liq. Cryst.* **574**, 60 (2013).

[43]   B. S. Scheuble, G. Weber, and R. Eidenschink, *Proc. Eurodisplay 84, Paris*, 65 (1984).

[44]   U. Krüger, R. Pepperl, and U. J. Schmidt, *Proc. IEEE* **61**, 992 (1973).

[45]   M. Beevers and G. Khanarian, *Aust. J. Chem.* **32**, 263 (1979).

[46]  J. M. Neto and A. B. Villaverde, *J. Phys. Condens. Matter* **8**, 2791 (1996).

[47]  P. G. de Gennes, *Phys. Lett. A* **30**, 454 (1969).

[48]  A. R. Johnston, *J. Appl. Phys.* **44**, 2971 (1973).

[49]  J. C. Filippini and Y. Poggi, *J. Phys. Lett.* **35**, 99 (1974).

[50]  M. Schadt, *J. Chem. Phys.* **67**, 210 (1977).

[51]  L. Schneider and J. H. Wendorff, *Liq. Cryst.* **22**, 29 (1997).

[52]  H. Khoshsima, H. Tajalli, A. G. Gilani, and R. Dabrowski, *J. Phys. D* **39**, 1495 (2006).

[53]  G. R. Luckhurst, *Thin Solid Films* **393**, 40 (2001).